\documentclass[12pt]{article}

\usepackage{newtxtext,newtxmath}

\usepackage{graphicx}

\usepackage[letterpaper,margin=1in]{geometry}

\renewenvironment{abstract}
	{\quotation}
	{\endquotation}

\date{}

\makeatletter
\renewcommand{\fnum@figure}{\textbf{Figure \thefigure}}
\renewcommand{\fnum@table}{\textbf{Table \thetable}}
\makeatother

\usepackage{scicite}

\usepackage{url}

\usepackage{textcomp}

\def\scititle{
Constraining an exoplanet's magnetic field using star-planet interactions
}
\title{\bfseries \boldmath \scititle}

\author{
	D.~Revilla$^{1\ast}$,
	P.J~Amado$^{1,\dagger}$,
	R.~Luque$^{1,2,\dagger}$,
    P.~Schöfer$^{1,\ddagger}$,
    A.F.~Lanza$^{3}$, \and
    A.~Binnenfeld$^{4,\S}$,
    J.A.~Caballero$^{5}$,
    A.P.~Hatzes$^{6}$,
    G.W.~Henry$^{7}$,
    S.V.~Jeffers$^{6}$,
    S.~Kaur$^{8,9}$,\and
    E.~Pallé$^{10,11}$,
    L.~Peña-Moñino$^{1}$,
    M.~Pérez-Torres$^{1,12}$,
    A.~Quirrenbach$^{13}$,\and
    A.~Reiners$^{14}$,
    I.~Ribas$^{8,9}$,
    D.~Viganò$^{8,9,15}$
    M.R.~Zapatero Osorio$^{5}$,
    S.~Zucker$^{4,16}$\and
	\small$^{1}$Instituto de Astrofísica de Andalucía - Consejo Superior de Investigaciones Científicas, Granada, Spain.\and
	\small$^{2}$Department of Astronomy and Astrophysics, University of Chicago, Chicago, USA.\and
    \small$^{3}$Osservatorio Astrofisico di Catania, Istituto Nazionale di Astrofisica, Catania, Italy.\and
    \small$^{4}$Porter School of the Environment and Earth Sciences, Raymond and Beverly Sackler Faculty of Exact Sciences, Tel Aviv University, Tel Aviv, Israel \and
    \small$^{5}$Centro de Astrobiología, Consejo Superior de Investigaciones Científicas - Instituto Nacional de Técnica Aeroespacial, Centro Europeo de Astronomía Espacial Campus, Villanueva de la Cañada, Madrid, Spain.\and
    \small$^{6}$Thüringer Landessternwarte Tautenburg, Tautenburg, Germany.\and
    \small$^{7}$Tennessee State University, Nashville, USA.\and
    \small$^{8}$Institut d’Estudis Espacials de Catalunya, Castelldefels, Barcelona, Spain.\and
    \small$^{9}$Institut de Ciències de I'Espai - Consejo Superior de Investigaciones Científicas, Campus Universidad Autónoma de Barcelona, Cerdanyola del Vallès, Barcelona, Spain.\and
    \small$^{10}$Instituto de Astrofísica de Canarias, La Laguna, Tenerife, Spain.\and
    \small$^{11}$Departamento de Astrofísica, Universidad de La Laguna, La Laguna, Tenerife, Spain.\and
    \small$^{12}$School of Sciences, European University Cyprus, Nicosia, Engomi, Cyprus.\and
    \small$^{13}$Landessternwarte, Zentrum für Astronomie der Universität Heidelberg, Heidelberg, Germany.\and
    \small$^{14}$Institut für Astrophysik und Geophysik, Georg-August-Universität, Göttingen, Germany.\and
    \small$^{15}$Institute of Applied Computing Community Code, Universitat de les Illes Balears, Palma, Spain.\and
    \small$^{16}$School of Physics and Astronomy, Raymond and Beverly Sackler Faculty of Exact Sciences, Tel Aviv University, Tel Aviv, Israel.\and
	\small$^\ast$Corresponding author. Email: drevilla@iaa.csic.es\and
	\small$^\dagger$These authors contributed equally to this work.\and
    \small$^\ddagger$ Present address: Department of Physics, Ariel University, Ariel, Israel.\and
    \small$^\S$ Present address: Laboratory of Astrophysics, Ecole Polytechnique Fédérale de Lausanne, Observatoire de Sauverny, Versoix, Switzerland.\and
}

\begin{document} 

\maketitle

\begin{abstract} \bfseries \boldmath

Theory predicts that a planet with a sufficiently strong magnetic field orbiting close to its host star could induce star-planet magnetic interactions. This is potentially observable as an optical or radio signal synchronised with the orbital period. We analyze 18 years of high-resolution optical spectroscopy of GJ 436, a low mass star orbited by a Neptune-sized exoplanet in a polar eccentric orbit. Stellar activity indicators show enhancements at a period corresponding to the exoplanet orbit, modulated by stellar rotation, and the star's 8-year magnetic cycle. We interpret this as a signal of star-planet magnetic interaction. Using a geometric model, we reproduce these periods if GJ 436 b has a magnetic field strength of 6 to 110 Gauss.

\end{abstract}

In a planetary system, the host star and its orbiting planets undergo continuous physical interactions, which can be gravitational, radiative, and magnetic. These interactions are typically strongly asymmetric, with the star dominating the planet. However, theory predicts this balance can shift in systems where the planet orbits very close to its host star. At small orbital distances, the planet could influence the star, producing measurable changes in stellar rotation, angular momentum evolution, or magnetic activity. These effects could arise from mechanisms including tidal interactions, spin-orbit coupling, or magnetic connections, which are predicted to have distinct observational signatures \cite{Lanza_2008, Strugarek_2014, Strugarek-Shkolnik_2025}. \\

Among these mechanisms, magnetic star-planet interaction (SPI) occurs when a planet interacts with a magnetized stellar wind. If the planet has a sufficiently strong magnetic field, this interaction is potentially observable and, under specific model assumptions, could constrain the strength of the planetary magnetic field \cite{Cauley_2019}. The magnetic field of a planet can determine whether it retains an atmosphere (and for how long), constrains its interior structure, and influences its long-term evolution \cite{Strugarek_2019, Vidotto_2013}. Similar magnetic coupling has been observed in the Solar System, such as the effect on Jupiter's magnetic field produced by its moons (Io and Ganymede) \cite{Prange_1996}.
Theoretical models predict that SPI could produce periodic signals that are potentially  observable at radio and optical wavelengths. Such signals would typically appear at the planet’s orbital period, but could potentially be shifted due to stellar rotation \cite{Shkolnik_2005}. The predicted strength of these signals depends on: the magnetic fields of the planet and the star, the planet's size, the orbital distance, and whether the planet is within the star’s Alfvén surface, the boundary within which magnetic forces dominate over kinetic forces \cite{Zarka_2018, Strugarek_2015}. Planetary systems with large planets on close orbits are therefore the most likely to exhibit SPI and there is evidence of SPI in several such systems \cite{Shkolnik_2003,Cauley_2018,Ilin_2025}.\\ 

\paragraph*{The GJ 436 system}
The red~dwarf star GJ~436 (Ross 905) (at a distance $d=9.78\,$ parsecs \cite{GaiaDR3}), has a radius of $R=0.42$ solar radii and mass $M=0.44$ solar masses \cite{Rosenthal_2021}. It has a rotation period of $P_{\rm rot}=45\pm5 \, \textrm{d}$ \cite{Bourrier_2018}. It is orbited by a large exoplanet, GJ~436~b \cite{Butler_2004}, which has a mass of $4.170 \pm 0.168$ Earth masses ($M_\oplus$) \cite{Maciejewski_2014} with an orbital period of $P_{\rm orb} = 2.64 \,d$  \cite{Butler_2004}. The stellar magnetic environment is predicted to be conducive to SPI \cite{Bourrier_2016,Lavie_2017}. The star is magnetically quiet \cite{Kumar_2022}, with low background variability in its chromosphere, the thin, hot layer of the stellar atmosphere just above the visible surface where magnetic activity manifests as excess emission from the ultraviolet to the near-infrared variability.\\

\paragraph*{Spectroscopic datasets}
Since the identification of GJ 436 b, to constrain its mass this system has been regularly observed using high-resolution optical spectroscopy. We re-analyze those archival observations to search for periodic signatures of SPI. We focused on chromospheric activity indicators predicted to be sensitive to SPI \cite{Cuntz_2000, Shkolnik_2003}. We sought to measure the Ca~{\sc ii}~{\textsc{H\&K}} lines (396.84\,nm and 393.36\,nm); the Ca~{\sc ii} infrared triplet at 849.8\,nm (\textsc{IRT-a}), 854.2\,nm (\textsc{IRT-b}) and 866.2\,nm (\textsc{IRT-c}); and H$\alpha$ at 656.28\,nm. We adopted archival high-resolution spectra observed with two instruments: the High Accuracy Radial velocity Planet Searcher (HARPS) \cite{Mayor_2003}, covering 380 to 690\,nm with a resolving power of 115,000 and Calar Alto high-Resolution search for M dwarfs with Exoearths with Near-infrared and optical Echelle Spectrographs (CARMENES) \cite{Quirrenbach_2014}, covering 520 to 1710\,nm, with a resolving power of 94,600 in the optical and 80,400 in the near-infrared channel. We analyzed a total of  371 spectra of GJ~436: 169 HARPS observations between 2006 and 2010, a further 23 HARPS observations in 2020, and 112 CARMENES observations in 2016, and 51 CARMENES observations in 2024.
As SPI signals are expected to be transient \cite{Shkolnik_2008, methods}, we applied time-series analysis techniques optimized to find periodic and non-continuous signals in unevenly sampled data \cite{methods}. We first used the generalized Lomb-Scargle (GLS) periodogram \cite{Ferraz-Mello_1981, Zechmeister_2009} to search for signals across a total of seven different epochs (5 from HARPS and 2 from CARMENES (Figs \ref{HARPS_2007full_All}-\ref{CARMENES_2024_All})), defined here as yearly observing intervals. Then, to account for the transitory nature of SPI, we applied rolling periodograms \cite{Herbort_2018, Schofer_2019} to each epoch to test whether the signal changes in intensity over shorter time scales. We define these as sub-epochs, corresponding to continuous observing runs spanning days to weeks within a given year. Then, once those sub-epochs are identified, we again applied to them the GLS periodogram to better characterize the properties of the modulation.\\

\begin{figure}
    \centering
    \includegraphics[width=0.8\textwidth]{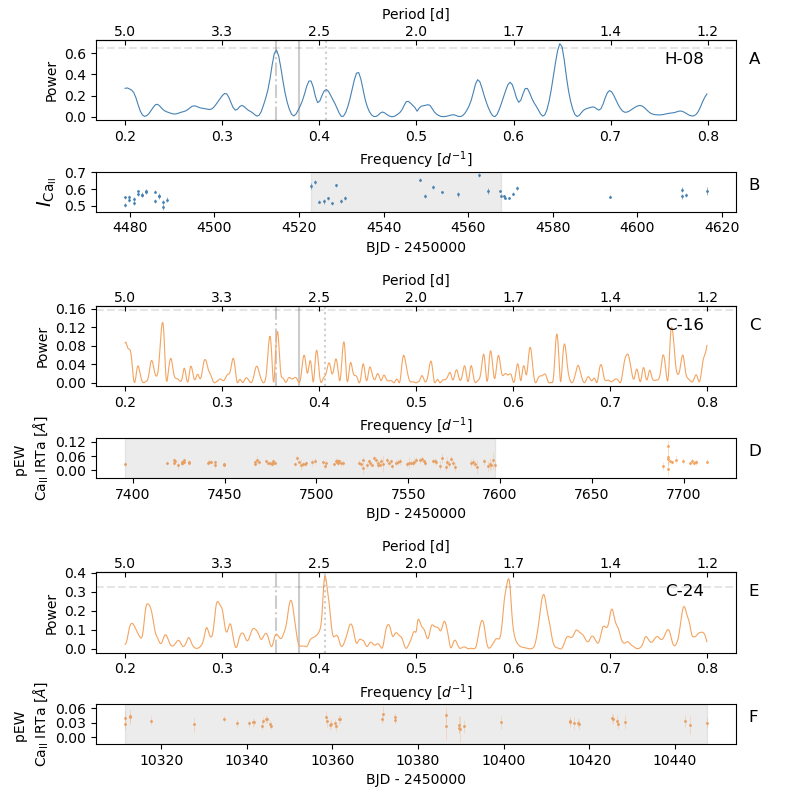}
    \caption{\textbf{Periodograms of calcium chromospheric indicators}. (\textbf{A}) GLS periodogram of the $I_{\text{Ca}_{\textsc{ii}}}$ index from the H-08 sub-epoch shown as a shaded gray region in panel B. Vertical lines mark the expected SPI periods at $P_{\text{syn}}=2.81 \,\,\text{d}$ (dot-dashed), a solid line at $P_{\rm orb} = 2.64 \,\,\text{d}$ (solid), and $P_{\text{a-syn}}=2.46 \,\,\text{d}$ (dotted). The dashed horizontal line shows 1\% FAP \cite{Zechmeister_2009}. (\textbf{B}) The $I_{\text{Ca}_{\textsc{ii}}}$ activity indicator time series of the 2008 HARPS epoch. Gray shading indicates the data used to calculate the periodogram in panel A. (\textbf{C} and \textbf{D}) Same as panels A and B, but for the pEW’(Ca~{\sc ii}~{\textsc{IRT-a}}) index in the C-16 epoch. (\textbf{E} and \textbf{F}) Same as panels C and D, but for the C-24 epoch. Each panel is color-coded, with blue indicating HARPS data and orange indicating CARMENES data.}
    \label{Fig_1}
\end{figure}

\paragraph*{Periodic stellar activity}
In the computed periodograms, we identified a peak at 2.81\,d — along with its 1-day alias at 1.54\,d — in the HARPS observations taken in 2008 (abbreviated H-08) and CARMENES observations taken in 2016 (C-16) (Fig. \ref{Fig_1}~A-B). This peak coincides with the system's synodic period, defined as $P_{\text{syn}} = \left(P_{\text{orb}}^{-1} - P_{\text{rot}}^{-1}\right)^{-1}$, which is one of the two possible periods at which SPI signatures have been predicted, the other being the orbital period \cite{Fischer_2019, Saur_2018}. 
The local False Alarm Probability (FAP) of the 2.81~d peaks appearing at that expected period \cite{Hatzes_2019} are 0.37\% in H08 and 0.77\% in  C-16. These FAP values were computed at a fixed pre-defined frequency, so represent the probability of a noise peak at that location in a single epoch, not anywhere in the periodogram at any epoch. Figure \ref{Fig_1} shows the global FAP values, which consider the entire frequency range. 

For the 2024 CARMENES epoch (C-24), no peak appears at 2.81\,d, with local FAP of 99.8\%, indicating no signal at the expected frequency. However, there is a peak at 2.46\,d, which has a global FAP of 0.17\% (Fig.~\ref{Fig_1}E).
The frequency of this second peak is consistent with a different expected period which we have called anti-synodic: $P_{\text{a-syn}} = \left(P_{\text{orb}}^{-1} + P_{\text{rot}}^{-1}\right)^{-1}$.
The C-16 periodogram shows an isolated peak at 4.19\,d (Fig.\ref{CARMENES_2016_All}), which does not coincide with any expected SPI signal. We were unable to determine the physical origin of this peak.\\

Each of these signals is detected in either the Ca~{\sc ii}~{\textsc{IRT-a}} or Ca~{\sc ii}~{\textsc{H\&K}} lines, which are known to be strongly correlated with each other \cite{Martin_2017} and are tracers of chromospheric activity \cite{Wilson_1968, Andretta_2005}. No epoch was observed with both instruments at the same time, so we were not able to compare the simultaneous behavior of both lines contemporaneously. We quantify these lines using the pseudo-equivalent width ( pEW') for Ca~{\sc ii}~{\textsc{IRT-a}} \cite{Schofer_2019} in the CARMENES spectra and $I_{\rm CaII}$ \cite{Kumar_2022} for the Ca~{\sc ii}~{\textsc{H\&K}} lines in the HARPS spectra. We also searched other activity indicators, arising from the star's chromosphere or photosphere, the photosphere being the visible surface of the star lying just below the chromosphere, including the He $\rm D_3$, Na D, H$\alpha$, TiO, VO and Ca~{\sc i} lines \cite{methods}, but found no periodicities. This indicates that the modulation originates in the lower chromosphere where Ca~{\sc ii} is present; for this type of star, the temperature of the lower chromosphere is $\sim$3000 to 6000\,K and the column mass density is (3 to 8) $\times 10^{-4}$\,kg\,m$^{-2}$ \cite{Hintz_2023}. H$\alpha$, which does not exhibit the 2.81\,d or 2.46\,d periodicities, forms higher in the stellar atmosphere at temperatures of 8000 to 11000\,K and lower column mass densities ($\sim10^{-4}$\,Kg\,m$^{-2}$).\\

\paragraph*{Geometric Model}
To investigate the origin of multiple periodicities in the data, we consider a geometric model containing two distinct activity regions: one associated with intrinsic stellar activity, and another induced by magnetic star–planet interaction (Figure \ref{fig:Coordinate}). The first region is fixed on the stellar surface and co-rotates with the star, while the second is magnetically linked to the planet and follows its orbital motion. This model accounts for the stellar rotation period, the planet's orbital period, and the obliquity of the planetary orbit \cite{methods}. The relative position between these two regions changes over time, producing a periodically varying separation that modulates the amplitude of the chromospheric emission.\\

\begin{figure}
    \centering
    \includegraphics[width=0.9\textwidth]{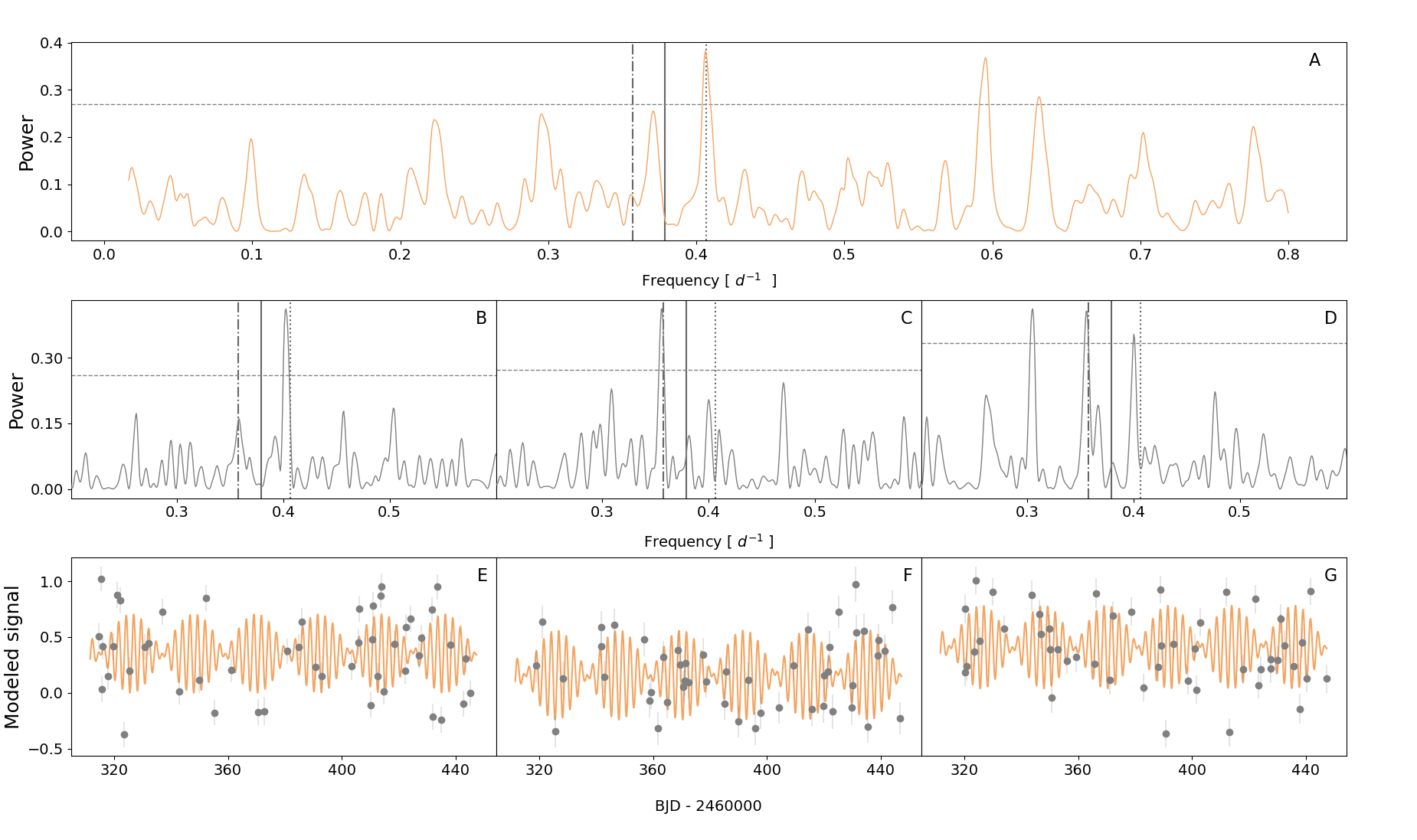}
    \caption{\textbf{Observations during C-24 compared to simulations.} \textbf{(A):} GLS periodogram of the pEW'(Ca~{\sc ii}~{\textsc{IRT-a}}) index during C-24. Line styles are the same as Fig. \ref{Fig_1}. \textbf{(B-D):} GLS periodogram of three synthetic datasets \cite{methods} to show how different mock observations following the geometric model result in the appearance of peaks at the two periods found in the periodogram of the observations. Panel B reproduces the 2.81\,d peak from C-16 and H-08, panel C reproduces the 2.46\,d peak from C-24, and panel D has several peaks. Line styles are the same as in Fig. \ref{Fig_1}. \textbf{(E-G):} The orange line is the simulated signal from the geometric model. Data points are synthetic observations at different times than the C-24 observations to show how the sampling affects the periodogram shown in panels B-D. We also added Gaussian-distributed noise with a standard deviation of $1\sigma$ of the modeled signal. Error bars show synthetic $1\sigma$ uncertainties generated derived from fitting a Gaussian distribution to the C-24 observations.}
    \label{Fig_2}
\end{figure}

Previous classical SPI models predicted periodic signatures at either the synodic or orbital periods \cite{Cuntz_2000,Lanza_2012,Fischer_2019}. We used simulations to generate synthetic observational datasets using our geometric model \cite{methods}. We find that only the synodic and anti-synodic frequencies appear, due to the system’s geometry and visibility conditions (Eq. \ref{dist2} and Fig. \ref{fig:Coordinate}). The synodic period $P_{\text{syn}}$ dominates if the planet's orbit is prograde (with the inclination being $I \ll 90^{\text{o}}$). If the planet is retrograde ($180^{\text{o}}>I \gg 90^{\text{o}}$) the anti-synodic $P_{\text{a-syn}}$ period dominates. In the intermediate case, for polar systems ($I\approx90^{\text{o}}$), in which the planet orbits perpendicular to the stellar equator, both periods appear with the same intensity. GJ~436~b has $I=103\pm13^{\text{o}}$ \cite{Bourrier_2018}, so we expect the latter regime to apply. The predicted signal exhibits a beat-like modulation composed of two periods (Fig.~\ref{Fig_2} E-F). The relative detectability of $P_{\text{syn}}$ and $P_{\text{a-syn}}$ depends on the observational cadence and signal-to-noise characteristics, so either, or both, may appear in a given dataset ( Fig.~\ref{Fig_2}~D and \cite{methods}). Our model does not predict a detectable signal at the orbital period $P_{\rm orb}$ under any considered conditions, even in ideal, continuous observations.\\

\begin{figure}
    \centering
    \includegraphics[width=0.8\textwidth]{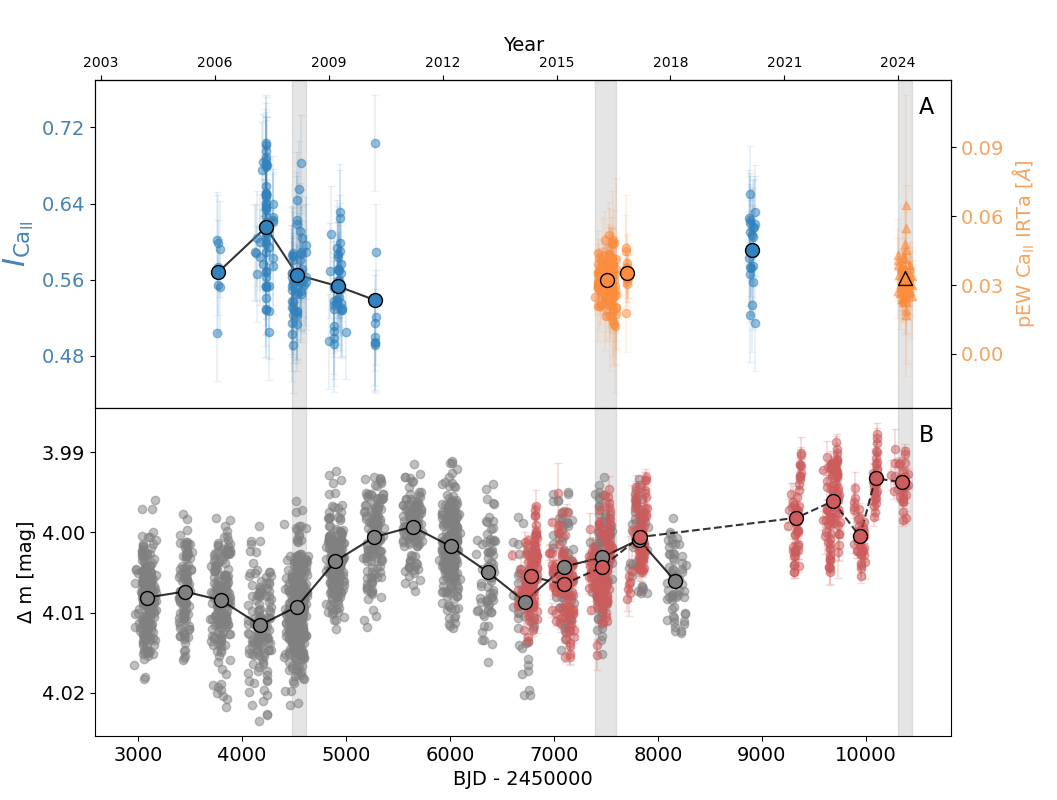}
    \caption{\textbf{Chromospheric activity indicators and photometric time series.} \textbf{(A)}: Chromospheric activity indicators. HARPS (blue) $I_{\text{Ca}_{\textsc{ii}}}$ on the left axis, and CARMENES (orange dots for C-16 and triangles for C-24)  pEW'(Ca~{\sc ii}~{\textsc{IRT-a}}) on the right axis. Filled circles are the yearly mean values in each epoch. \textbf{(B)}: Differential photometry from T12 APT combined Stromgren b- and y-band data (gray) and AIT (red) after correcting the offset of the difference calculated in the four contemporary epochs and plotted together. Filled circles represent the same as in panel A. The black and dashed lines connect the seasonal means of the T12 APT and C14 AIT datsets respectively. Gray shaded regions show the epochs where SPI was detected (H-08, C-16, and C-24).}
    \label{Fig_3}
\end{figure}

\paragraph*{Stellar activity cycle}
Our geometric model does not explain why SPI signals are only detected during specific epochs, in 2008, 2016, and 2024. Red~dwarf stars are known to exhibit magnetic activity cycles on multi-year timescales, which manifest in both photometric variability and chromospheric emission indicators \cite{Robertson_2013, Suarez-Mascareño_2016}. Previous studies have determined that GJ 436's activity cycle is $7.75 \pm 0.10$ years  \cite{Lothringer_2018, Loyd_2023}. We analyzed public and unpublished observations of GJ 436 with the Tennessee State University's 0.8 m Automatic Photoelectric Telescope (APT) from 2003 to 2018 and the Celestron 14-inch Automated Imaging Telescope (AIT) from 2013 to 2024 \cite{methods}. We measured photometry from those observations and compared it to the chromospheric calcium indices from HARPS and CARMENES (Fig.~\ref{Fig_3}). Although the AIT data appear less regular than in previous observations, they remain consistent with an approximately 8-year cycle. The three epochs in which we detect SPI signals are spaced about 8 years apart, coinciding with intermediate activity levels.

We suggest that this temporal coincidence could arise if the detectability of SPI signals is modulated by the stellar magnetic cycle. The total power involved in magnetic interaction depends on the star’s magnetic field strength \cite{methods}. Although stronger magnetic fields at the maximum of the activity cycle could enhance SPI power, the concurrent increase in the star's own activity, with more and larger actively magnetic regions, could overwhelm the localized signal produced by SPI \cite{Hathaway_2015}. Conversely, during activity minima, the overall magnetic energy available for SPI is reduced, making it more difficult to detect. We propose that intermediate activity levels —  when the field is strong enough to support SPI, but surface activity is relatively low, offer more favorable conditions for identifying SPI-induced variability.\\

The detection of signals at two distinct periods is consistent with our geometric model, and their re-appearance at multiple epochs separated by approximately eight years, indicates an SPI origin. We quantified the statistical significance by computing the FAP of the 2.81~d peaks in the H-08 and C-16 datasets, and of the 2.46~d peak in the C-24 data. Standard 
analytical FAPs work by searching over a broad range of frequencies, 
but our peaks are expected at specific values, the planetary orbital period, or a combination of that with the stellar rotation 
period. We therefore used a bootstrap procedure that incorporates this prior knowledge 
\cite{methods}: the data points are randomly shuffled many times to simulate pure 
noise, and the FAP is estimated from how often a noise realization produces a peak 
as strong as the observed one near the predicted frequency. The resulting FAPs 
are $3.7 \times 10^{-3}$ for H-08, $7.7 \times 10^{-3}$ for C-16, and 
$3.2 \times 10^{-4}$ for C-24 (Table~S1), meaning that random noise would produce 
peaks this strong only in 0.4\%, 0.8\%, and 0.03\% of cases, respectively.

For an overall FAP that also accounts for the four non-detection epochs, we applied Fisher's test \cite{Fisher_1925}.
This method sums the logarithms of the per-epoch FAPs 
into a single statistic, $X^2 = -2 \sum_i \ln p_i$ (where $p_i$ are the individual 
FAPs), and compares it to the distribution expected if all seven epochs were pure 
noise. Strong detections produce large contributions to $X^2$, while epochs without 
signal add only small terms, so the four non-detections appropriately dilute the 
combined statistic rather than being ignored. The seven epochs are statistically 
independent: they are separated by years and any correlated structure within each epoch is removed by the shuffling procedure. 
Applying this test to all seven epochs gives $X^2 = 33.78$ (Fig. \ref{fig:Fischer}), corresponding to an 
overall probability of approximately $2.22 \times 10^{-3}$ that the observed pattern 
of detections and non-detections could have arisen from noise alone. This means that we reject the null hypothesis of no SPI signal at the 99.7\% confidence level.
We interpret them as arising from a common recurring physical mechanism.\\

To produce magnetic SPI, the planet must be within the stellar Alfvén surface. Spectropolarimetric observations in 2016 \cite{Kumar_2022} have previously been used to reconstruct GJ~436’s large-scale magnetic field and determine the location of its Alfvén surface \cite{Bellotti_2023, Vidotto_2023}. This surface expands approximately between $\sim 1.5$ and $\sim 3$ times the orbital distance of the planet (14.56 $R_\star$ $\simeq 0.028$ au). This indicates that the planet was located within the Alfvén surface for most of its orbit during the C-16 epoch. Although the Alfvén surface can change during the stellar activity cycle, the other SPI signals were detected during the H-08 and C-24 epochs, which are separated from 2016 by nearly one stellar activity cycle. We therefore find it is likely that GJ~436~b was also within the Alfvén surface during those epochs.\\

\paragraph*{Planetary Magnetic Field}
The power emitted in the Ca~{\sc ii}~{\textsc{IRT-a}} line during the 2016 and 2024 epochs was $10^{19}\,\text{W}$ \cite{methods}, calculated using the observed line strengths into absolute fluxes using empirical calibrations that relate the brightness of the line relative to the star's total luminosity \cite{Labarga_2022}. However, this is only a fraction of the total SPI power, because most is likely converted to heat or emitted at other wavelengths. Because the spectral energy distribution of SPI events is unknown, we adopt a flare observed on the red~dwarf AD Leo as an empirical proxy for a heated chromosphere. This flare radiated about 2 percent of its energy in the Ca~{\sc ii}~{\textsc{IRT-a}} line and $< 50$ percent in the Ca~{\sc ii}~{\textsc{H\&K}} lines. We therefore estimate that the Ca~{\sc ii}~{\textsc{IRT-a}} line accounts for 2 to 100 percent of the total SPI power.\\

To convert the observed SPI power into constraints on the planetary magnetic field strength ($B_p$) and planet magnetospheric radius ($r_M$), we evaluated several theoretical models of magnetic interaction: the Alfvén wing model, the magnetic reconnection model, the helicity variation model, and the interconnecting magnetic loop model \cite{methods}. The magnetospheric radius defines the distance at which the planetary field balances the stellar wind pressure. A larger magnetosphere increases the area over which the stellar wind interacts with the planetary field, thereby enhancing the total power that can be extracted from the system. Among the models considered, only the theoretical model that contemplates an interconnecting magnetic loop between the planet and the star \cite{Lanza_2013} produced energy outputs consistent with our observations. This model assumes a stable magnetic configuration in which a magnetic-loop-like structure connects the star and planet; orbital motion stresses this structure and injects energy that is rapidly dissipated to maintain equilibrium. The other models we considered do not reproduce the observed power levels. Table \ref{Tab:Bfields} presents the derived values of $B_p$ and $r_M$ using three assumed stellar magnetic field strengths in gauss (G) at the planet’s orbital distance: 0.013~G, 0.026~G (\cite{Vidotto_2023}, their Models I and II) and 0.13~G, \cite{Lanza_2018}. Each of these values is consistent with the observed power, but different SPI emission efficiencies.\\

The planetary magnetic field strength required to reproduce the observed Ca~{\sc ii}~{\textsc{IRT-a}} emission depends on both the stellar magnetic field at the orbital distance and the fraction of the total interaction power emitted in this line. Using the interconnecting loop model, we derive a lower bound of 6.3\,G by assuming a radial stellar field (0.13\,G at the planet distance) and that 100 percent of the SPI power is radiated in Ca~{\sc ii}~{\textsc{IRT-a}}. At the other extreme, if only 2 percent of the power is emitted in this line and the stellar field, is weaker (0.013\,G \cite{Vidotto_2023}), the required planetary field is 110\,G, which we regard as the upper bound. This range reflects the uncertainty in the total power dissipated by the interaction, the variation in assumed stellar magnetic field strength and the model’s dependence on these parameters.

The corresponding magnetospheric radii are 6 to 21 planetary radii ($R_p$); for comparison Neptune’s magnetosphere is approximately 26 planetary radii \cite{Ness_1989}. This difference could be due to GJ~436~b’s proximity to its host star and the stronger stellar wind pressure it experiences \cite{Vidotto_2013}. Our transient detections of SPI indicate that long-term variability in stellar magnetic topology and field strength can modulate the detectability of planetary magnetic interactions.\\


\clearpage


\begin{table} 
	\centering
	\caption{\textbf{Planetary magnetic field and magnetospheric radius of GJ~436~b.} The calculated planetary magnetic field strength ($B_p$, in gauss) and magnetospheric radius ($r_M$, in planetary radii) in the interconnecting loop model, under different assumptions about stellar magnetic field strength and SPI emission efficiency. Scenarios A and B use two extrapolations of the stellar field (\cite{Vidotto_2023} their Models I and II). Scenario C assumes a purely radial stellar field extrapolation \cite{Lanza_2018}. The percentages are the assumptions of the fraction of energy converted to flux in the Ca~{\sc ii}~{\textsc{IRT-a}} or Ca~{\sc ii}~{\textsc{H\&K}} lines. All scenarios assume a planetary radius of 4.2 Earth radii \cite{methods}.\\} 
	\label{Tab:Bfields} 
	
	\begin{tabular}{c | c c | c c | c c | c} 
	
		\hline
		\% Energy in Ca~{\sc ii} & 2 \% & & 25 \% & & 100 \% & & Ref. for $r_M$\\
		 Scenario & {$B_p$} (G) & {$r_M$} ($R_p$) & {$B_p$} (G) & {$r_M$} ($R_p$) & {$B_p$} (G) & {$r_M$} ($R_p$) \\
		\hline
		A & 110 & 20 & 24 & 12 & 10.3 & 9 &  Eq. \ref{Eq:rM_Vidotto} and Model I in \cite{Vidotto_2023} \\
		B & 95 & 13 & 21 & 8 & 9 & 6 &  Eq. \ref{Eq:rM_Vidotto} and Model II in \cite{Vidotto_2023}\\
        C & 68 & 21 & 14.7 & 13 & 6.3 & 9.6 &  Eq. \ref{Eq:rM_Lanza} \cite{Lanza_2018} \\
		\hline
	\end{tabular}
\end{table}


\clearpage 

%
\bibliography{Main_text} 
\bibliographystyle{sciencemag}


\section*{Acknowledgments}
We thank Giovanni M. Mirouh for discussion of the geometrical model. 
This publication is partly based on observations collected under the CARMENES guaranteed time observations and 24A-3.5-013 proposal.\\
CARMENES is an instrument at the Centro Astron\'omico Hispano en Andaluc\'ia (CAHA) at Calar Alto (Almer\'{\i}a, Spain), 
operated jointly by the Junta de Andaluc\'ia and the Instituto de Astrof\'isica de Andaluc\'ia (CSIC).
This instrument was also funded by the Complementary R\&D\&I Plan in Astrophysics and High-Energy Physics (AST\_00001\_8) with funding from the European Union – NextGenerationEU, the Spanish Government's Ministry of Science, Innovation and Universities, the Recovery and Resilience Plan, the Spanish National Research Council (CSIC), and the Andalusian Regional Government's Ministry of University, Research and Innovation
\paragraph*{Funding:}

 D.R. and L.P. were supported by the PRE2021-099654 and PRE2020-095421 grants, funded by MCIU/AEI/10.13039/501100011033. D.R. and P.A. acknowledge support from the Spanish National grant PID2022–137241NB–C42.
 P.A. acknowledge support from PID2022-137241NB-C43 project.
D.R., P.A., L.P. and M.P. acknowledge financial support from the Severo Ochoa grant CEX2021-001131-S funded by MCIN/AEI/ 10.13039/501100011033.
R. L. was supported by the European Union (ERC, THIRSTEE, 101164189) and NASA through the NASA Hubble Fellowship grant HST-HF2-51559.001-A awarded by the Space Telescope Science Institute, which is operated by the Association of Universities for Research in Astronomy, Inc., for NASA, under contract NAS5-26555.
I.R. was supported by the PID2024-158486OB-C31 MCIU CEX2020-001058-M MCIU "SPOTLESS" No. 101140786 Horizon Europe ERC founds.
G.H. was supported by NASA and NSF.
A.B. and S.Z. were supported by the Israel science foundation (grant No.\ 1404/22) and the Israel Ministry of Science and Technology (grant No.\ 3-18143).
S.K and D.V. acknowledge the support by the European Research Council (ERC) under the European Union’s Horizon 2020 research and innovation programme (ERC Starting Grant "IMAGINE" No. 948582). I.R. S.K. and D.V. were supported by a ``Mar\'ia de Maeztu'' award to the Institut de Ciències de l'Espai (CEX2020-001058-M). S.K. was supported by 
the doctoral program in Physics of the Universitat Autònoma de Barcelona. 
A.F.L. was supported by the Italian Ministry of University and Research through funding to the Italian National Institute for Astrophysics.
J.A.C. and M.R.Z.O. acknowledge financial support from PID2022-137241NB-C42 project.
I.R. acknowledge financial support from PID2024-158486OB-C31 MCIU, CEX2020-001058-M MCIU
and "SPOTLESS" No. 101140786  Horizon Europe ERC projects.
E.P. was supported by the European Union (ERC AdvG SPEAR, GA 101200674). Views and opinions expressed are however those of the authors only and do not necessarily reflect those of the European Union or the European Research Council. Neither the European Union nor the granting authority can be held responsible for them. E.P acknowledge financial support from Agencia Estatal de Investigaci\'on of the Ministerio de Ciencia e Innovaci\'on MCIN/AEI/10.13039/501100011033 and the ERDF “A way of making Europe” through projects PID2021-125627OB-C32 and PID2024-158486OB-C32.

\paragraph*{Author contributions:}
D.R. and P.A. conceived the study. D.R. led the analysis of the data. D.R., P.A and R.L interpreted the data. D.R., R.L. and P.A. wrote the manuscript. P.S. determined the CARMENES activity indicators and helped to interpret the data. A.F.L. developed the geometrical model. A.H. contributed to the statistical analysis. G.H. provided and analyzed the T12 APT and C14 AIT data. A.B. and S.Z. analyzed the spectra.
J.A.C., A.Q., A.R. and I.R. contributed to CARMENES operations, observations, and data extraction. S.V.J., S.K., E.P., L.P., M.P., D.V. and M.R.Z.O. contributed to the discussion and interpretation of the geometric model.

All authors provided comments on the manuscript.
\paragraph*{Competing interests:}
The authors declare no competing interests.
\paragraph*{Data and materials availability:}
The HARPS spectroscopic observations are available at: \url{http://archive.eso.org/eso/eso_archive_main.html} under the target name GJ 436. The CARMENES 2016 spectroscopic observations are available at: \url{http://carmenes.cab.inta-csic.es/gto/jsp/dr1Public.jsp.} under the names J11421+267 and Ross 905. The T12 APT and C14 AIT photometric observations are available at \cite{Revilla_2025data}. All the datasets used to produce the plots are also available at \cite{Revilla_2025data}.
No physical materials were generated in this work.


\subsection*{Supplementary materials}
Materials and Methods\\
Figs. S1 to S19\\
Table S1 to S2\\
References (51-85)  


\newpage


\renewcommand{\thefigure}{S\arabic{figure}}
\renewcommand{\thetable}{S\arabic{table}}
\renewcommand{\theequation}{S\arabic{equation}}
\renewcommand{\thepage}{S\arabic{page}}
\setcounter{figure}{0}
\setcounter{table}{0}
\setcounter{equation}{0}
\setcounter{page}{1} 


\begin{center}
\section*{Supplementary Materials for\\ \scititle}

	D.~Revilla, \and
	P.J~Amado, \and
	R.~Luque, \and
    P.~Schöfer, \and
    A.F.~Lanza, \and
    A.~Binnenfeld, \and
    J.A.~Caballero, \and
    A.P.~Hatzes, \and
    G.W.~Henry, \and
    S.V.~Jeffers, \and
    S.~Kaur,\and
    E.~Pallé, \and
    L.~Peña-Moñino, \and
    M.~Pérez-Torres, \and
    A.~Quirrenbach, \and
    A.~Reiners, \and
    I.~Ribas, \and
    D.~Viganò, \and
    M.R.~Zapatero Osorio, \and
    S.~Zucker\and \\
\small$^\ast$Corresponding author. Email: drevilla@iaa.csic.es\\
\end{center}

\subsubsection*{This PDF file includes:}
Materials and Methods\\
Figs. S1 to S19\\
Table S1 to S2\\
References

\newpage


\subsection*{Materials and Methods}

\subsubsection*{Observations and data reduction}
\paragraph{HARPS spectroscopy} \label{HARPS}
There are 192 observations of GJ 436 taken with the HARPS instrument \cite{Mayor_2003} at the 3.6m ESO telescope in La Silla, Chile, between January 2006 and March 2020. We followed previous methods \cite{Kumar_2022} for reducing the data and obtaining the activity indicators. The 2008 epoch consists of 47 observations. The HARPS spectra were not continuum-normalized \cite{Kumar_2022} and were used to derive magnetic activity indices for  Ca~{\sc ii}~{\textsc{H\&K}}, He $\rm D_3$, Na D and $H\alpha$. We also obtained the index for the Ca~{\sc i} (657.28 nm), a photospheric line that is not sensitive to chromospheric activity.\\

We use the \texttt{ACTIN2} package \cite{da_Silva_2011, da_Silva2018, da_Silva_2022} to calculate the indices as:
\begin{equation}
    I_{line} = \frac{\sum_{i=1}^{P} F_i}{\sum_{i=1}^{Q} R_j}
\end{equation}

where $F_i$ is the flux in the activity lines, $R_j$ the flux in the reference regions \cite{da_Silva2018,dasilva_2021}, while \textit{P} and \textit{Q} are the number of activity lines and reference regions used.

Because it consists of two lines, the Ca~{\sc ii}~{\textsc{H\&K}} index was calculated as:
\begin{equation}
    I_{\rm CaII} = \frac{F_{\rm Ca_{II}\,K} + F_{\rm Ca_{II}\,H}}{F_1 + F_2}
\end{equation}\\
where $F_{Ca_{II}K}$ and $F_{Ca_{II}H}$ represent the mean fluxes integrated in triangular windows around the K and H line cores and $F_1$ and $F_2$ are the mean fluxes in square windows of the pseudo-continuum. Calculation of all activity indices followed previous methods [\cite{Kumar_2022}, their section 3].

\paragraph{CARMENES spectroscopy} \label{CARMENES}
GJ 436 was observed with the CARMENES instrument \cite{Quirrenbach_2013,Quirrenbach_2014} at the 3.5\,m telescope of the Calar Alto Observatory in Almeria, Spain. There are 444 observations taken between January 2016 and April 2021, of which only 112 (with an average cadence of one observation every two days) were used for this study. The other observations were primarily high-cadence, intranight observations that were not used because they were sparse and not in continuous days. Additional observations of GJ~436 were obtained in 2024 (program ID: 24A-3.5-013) at the same phase of the stellar magnetic cycle as the SPI signals we identified in the H-08 and C-16 epochs. This additional program collected 51 observations from 2024 January 1 to May 16, with an average cadence of one observation every three days. Of the 51 observations, two were discarded due to low signal-to-noise ratio (those taken on Barycentric Julian Date (BJD) = 2460387.34146210 and 2460387.65357420). We derived pseudo-equivalent widths (pEW') of the He $\rm D_3$, Na D, H$\alpha$, and Ca~{\sc ii}~{\textsc{IRT}} lines (Ca~{\sc ii}~\textsc{IRT-a}, Ca~{\sc ii}~\textsc{IRT-b} and Ca~{\sc ii}~\textsc{IRT-c}) (with Ca~{\sc ii}~\textsc{IRT-a} being the bluest of the three) using a spectral subtraction technique and indices of selected TiO and VO absorption bands following previous methods \cite{Schofer_2019}. The wavelength ($\lambda$) intervals were $\Delta \lambda = 5\,\mathring{\mathrm{A}}$ for $H \alpha$,  $\Delta \lambda = 2.8\,\mathring{\mathrm{A}}$ for  Ca~{\sc ii}~{\textsc{IRT-a}} and b,  $\Delta \lambda = 2.4\,\mathring{\mathrm{A}}$ for Ca~{\sc ii}~{\textsc{IRT-c}},  $\Delta \lambda = 5.98\,\mathring{\mathrm{A}}$ for Na D and $\Delta \lambda = 2.5\,\mathring{\mathrm{A}}$ for He~$\text{D}_3$. We tested using a narrower interval but found that had a negligible effect.

\paragraph{T12 APT and C14 AIT photometry}
The Tennessee State University (TSU) 0.80\,m Automatic Photoelectric Telescope (APT) (hereafter T12 APT), located at the Fairborn Observatory in southern Arizona, is equipped with a two-channel photometer that uses two EMI 9124QB bi-alkali photomultiplier tubes to measure stellar brightness in the Str\"omgren $b$ and $y$ passbands simultaneously.
We use a total of 2040 photometric observations of GJ~436 taken during 15 observing seasons between 2003 and 2018. The first 14 observing seasons (from 2003 to 2017) have previously been presented \cite{Lothringer_2018}. The remaining seasons are published in this work. \\


The observations of GJ~436 were made differentially with respect to three nearby (on the sky) comparison stars:  HD~102555 , HD~103676, and HD 99518.  We measure the difference in brightness between GJ~436 and the three comparison stars and calculate differential magnitudes in the following six combinations:  GJ~436 $-$  HD~102555, GJ~436 $-$ HD~103676, GJ~436 $-$ HD 99518, HD 99518 $-$  HD~102555,  HD 99518 $-$  HD~103676, and HD~103676 $-$  HD~102555.  Intercomparison of these six light curves shows that comparison stars HD~102555 and HD~103676 both are apparently constant, so we present our results as differential magnitudes: GJ~436 minus the mean brightness of stars HD~102555 and HD~103676. To improve the photometric precision of the individual nightly observations, we combined the differential $b$-band and $y$-band magnitudes into a single $(b+y)/2$ passband.  The observations are all corrected for differential atmospheric extinction and transformed to the standard Str\"omgren $b$ and $y$ passbands using nightly observations of a set of standard stars distributed throughout the sky.\\

\begin{figure}
    \centering
    \includegraphics[width=0.9\columnwidth]{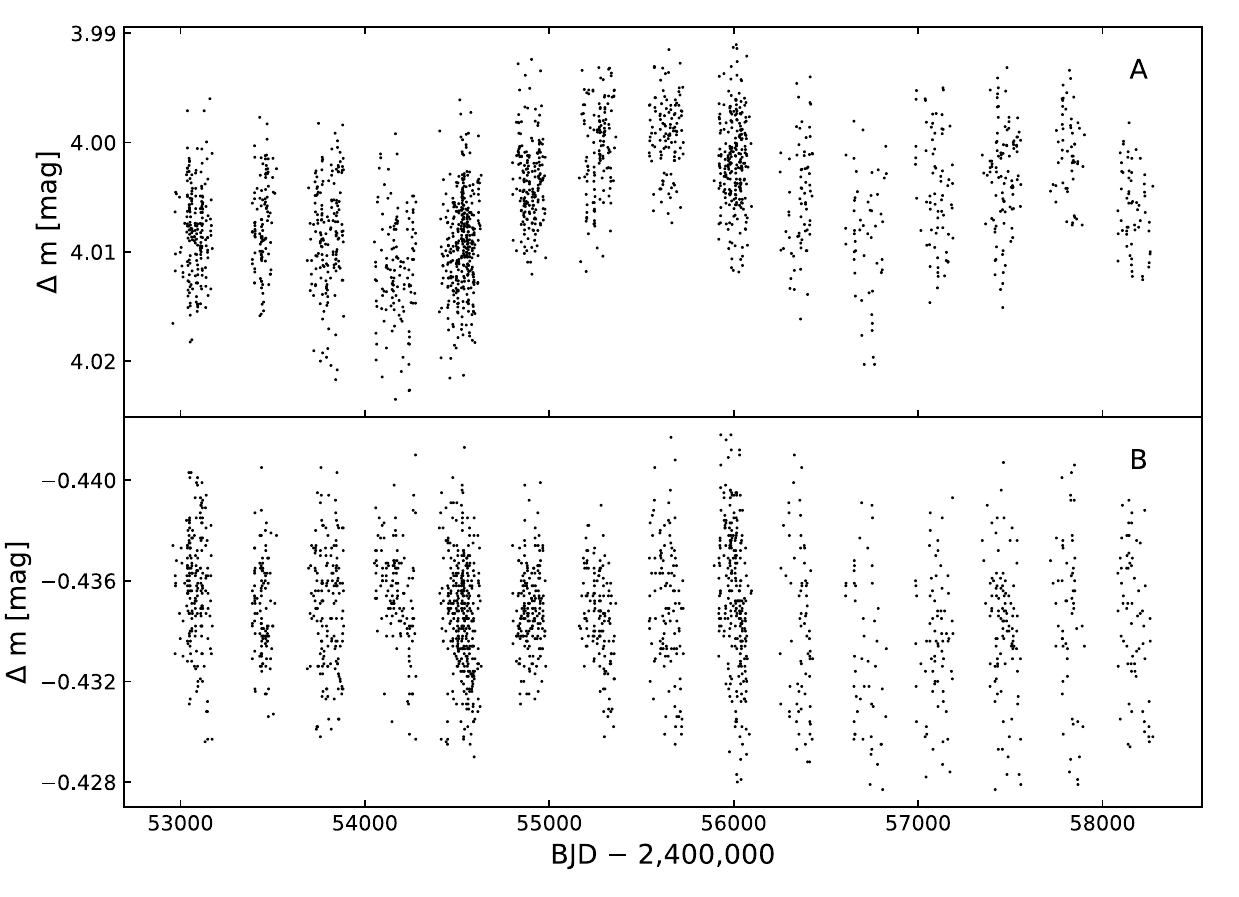}
    \caption{\textbf{T12 APT Photometry of GJ 436}. \textbf{(A):} Differential photometry from T12 APT combining Strömgren b and y-band data of GJ~436 compared the mean brightness of reference stars HD 102555 and HD 103676. \textbf{(B):} Differential magnitudes of the comparison stars, HD 102555 and HD 103676, at the same scale. A comparison of the two panels indicates low-amplitude variability in GJ 436 on both night-to-night and year-to-year timescales.}
    \label{Fig:APT-Greg}
\end{figure}

The T12 APT observations are shown in Fig. \ref{Fig:APT-Greg}, and the differential magnitudes of the two comparison stars in Fig. \ref{Fig:APT-Greg} B. The standard deviation of the 2043 nightly comparison star observations from their mean is 2.43 milli-magnitudes in the $(b+y)/2$ passband, which we interpret as the precision of the individual differential magnitudes. We find that GJ~436 exhibits low-amplitude night-to-night and year-to-year brightness variability.\\

We also acquired 607 additional observations in the Cousins $R$ band with the TSU Celestron 14-inch Automated Imaging Telescope (hereafter C14 AIT) covering  eight observing seasons: 2013 to 2017 and 2020 to 2024. The C14 AIT uses an SBIG STL-1001E Charge-Coupled Device (CCD) camera with a Kodak KAF-1001E detector, with a field of view of 21 $\times$ 21 arcmin.  Each nightly observation consisted of 4 to 10 consecutive exposures centered on GJ~436.  The individual nightly frames were co-added and reduced to differential magnitudes, calculated as GJ~436 minus the mean brightness of four constant comparison stars in the same field of view.  Each nightly observation was corrected for bias, ﬂat-ﬁelding, pier-side offset, and differential atmospheric extinction.\\

\begin{figure}
    \centering
    \includegraphics[width=0.9\columnwidth]{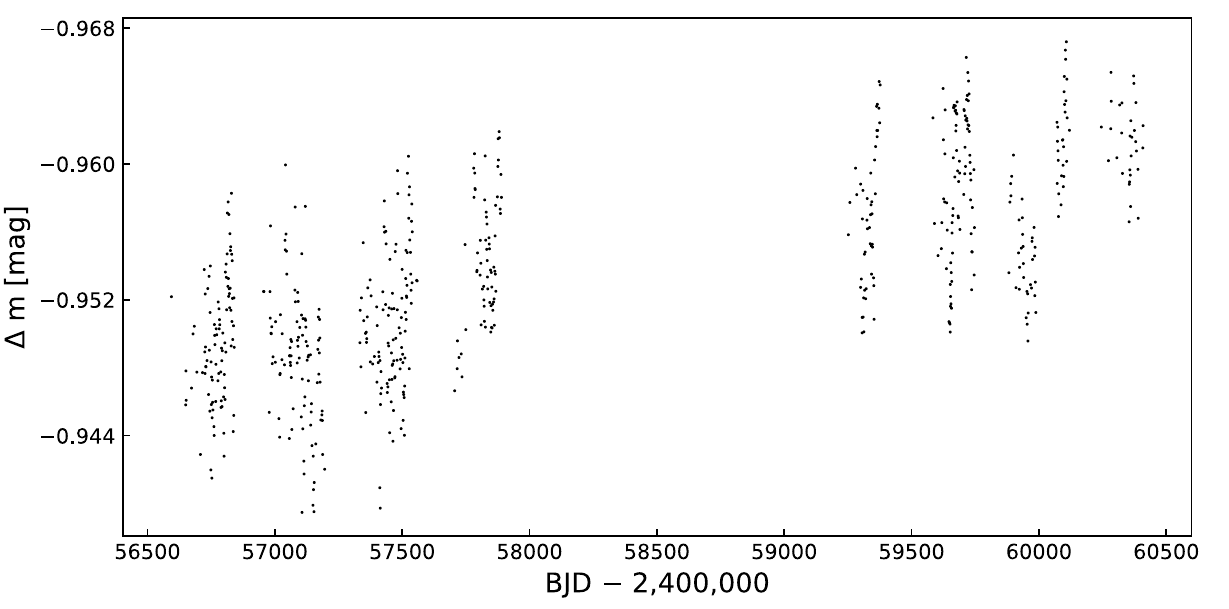}
    \caption{\textbf{C14 AIT Photometry of GJ 436}. Same as Fig. \ref{Fig:APT-Greg}, but showing the Cousins $R$ band differential photometry during the observing seasons 2013 to 2016 and 2020 to 2023. There is low-amplitude variability on both night-to-night and year-to-year timescales.}
    \label{Fig:C14-Greg}
\end{figure}

The C14 AIT observations are plotted in Fig. \ref{Fig:C14-Greg}. The first four observing seasons overlap with the T12 APT observations. The C14 AIT observations show the same low-amplitude nightly and yearly brightness variations as seen in the T12 APT data.\\

\subsubsection*{Periodograms and analysis procedures}
\paragraph{Splitting datasets}\label{subsub:GLSchunk}
There is theoretical uncertainty regarding when and for how long an SPI signal might manifest itself. This lack of knowledge complicates our search, because a signal might only be present at certain epochs. We therefore split the dataset into smaller chunks, which reduces the risk of diluting any transient signal that could otherwise be lost in the full dataset. This method allows us to isolate moments where the SPI signal is potentially detectable, focusing on smaller windows of time where the interaction might be more evident.\\

\begin{figure}
    \centering
    \includegraphics[width=\columnwidth]{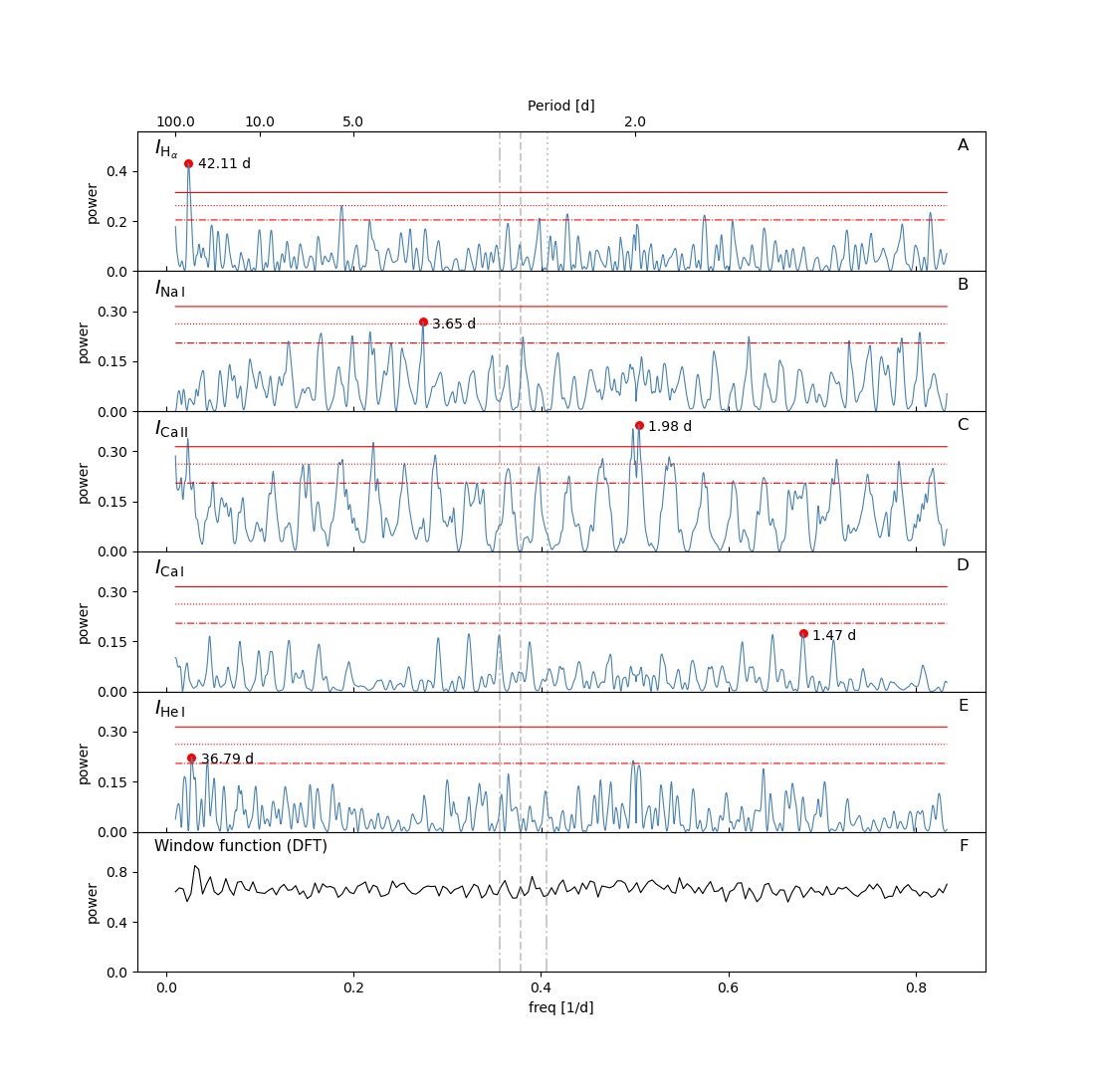}
    \caption{\textbf{GLS periodogram of the H-07 sub-epoch for all activity indicators measured using HARPS.} (\textbf{A-E}): The blue line shows the periodogram for each activity indicator, labeled in the left top corner of each panel. The three horizontal red lines mark the 10 (dashed), 1 (dotted) and 0.1\% (solid) FAP levels from the GLS, while the gray vertical ones mark the signals at 2.81\,d (dashed and dotted), 2.64\,d (dashed) and 2.46\,d (dotted). A red dot highlights the highest peak in each periodogram. (\textbf{F}): Periodogram of the window function.}
    \label{HARPS_2007full_All}
\end{figure}

\begin{figure}
    \centering
    \includegraphics[width=\columnwidth]{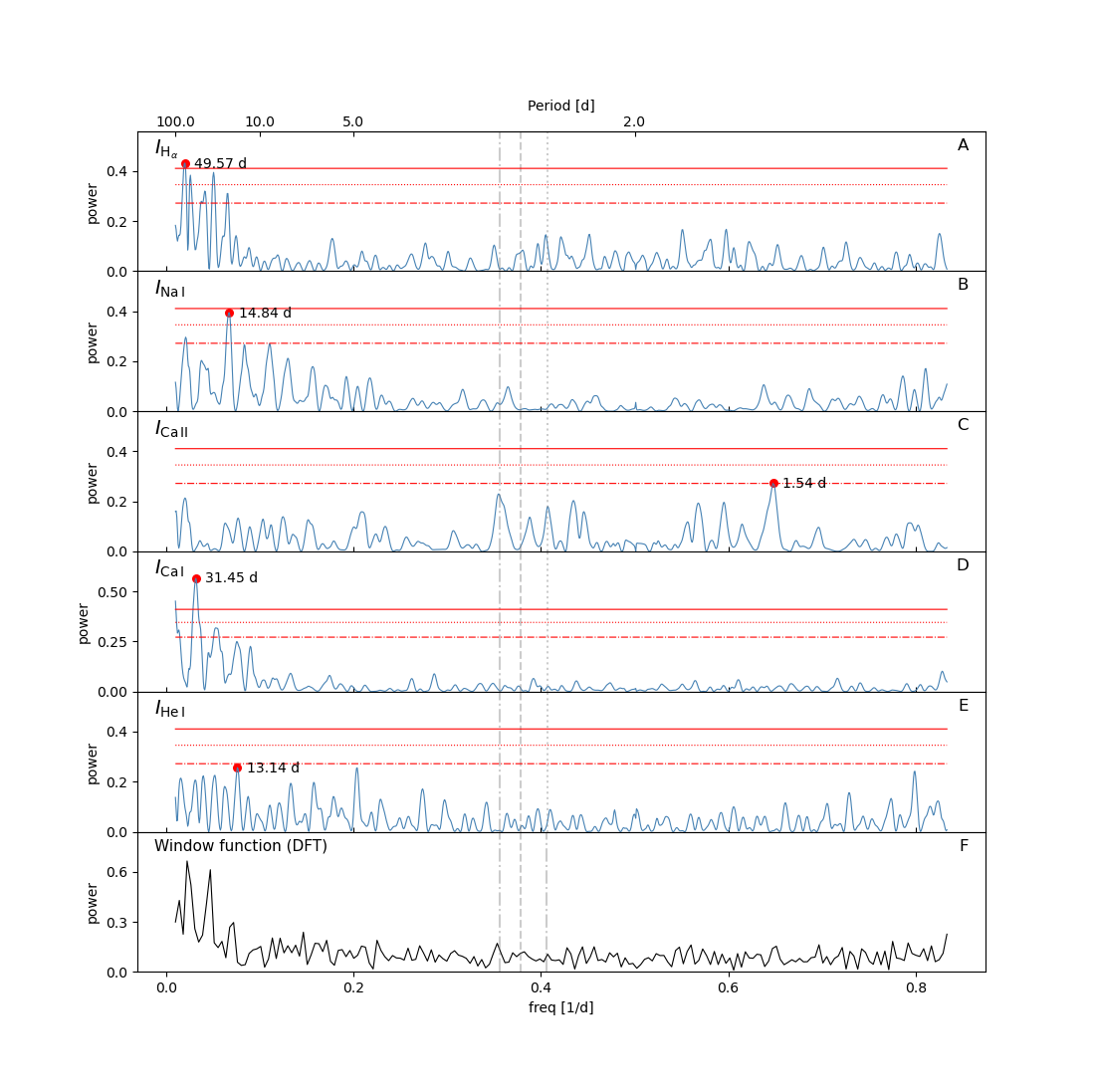}
    \caption{\textbf{Same as Fig.~\ref{HARPS_2007full_All}, but covering the entire 2008 observational season.}}
    \label{HARPS_2008full_All}
\end{figure}

First, we used the generalized Lomb-Scargle (GLS) periodogram \cite{Ferraz-Mello_1981, Zechmeister_2009} to search for signals across a total of seven different epochs (5 from HARPS and 2 from CARMENES), defined here as yearly observing intervals. The results are shown in Figs.~\ref{HARPS_2007full_All} through \ref{CARMENES_2024_All}. For HARPS, we divide the dataset in multiple epochs corresponding to the visibility of the target (January to June). Out of the six epochs and from all the activity indicators, we only detected a signal compatible with SPI at $P_{\text{syn}} = 2.81\,\text{d}$ in the Ca~{\sc ii}~{\textsc{H\&K}} index during 2008 (H-08). The global false alarm probability (FAP) of this signal is about 1\% (panel A in Fig. \ref{Fig_1}). For CARMENES, all the data were kept in two chunks corresponding to the 2016 (C-16) and 2024 (C-24) observing seasons. In the C-16 epoch, we find a peak at $P_{\text{syn}} = 2.81\,\text{d}$ in the Ca~{\sc ii}~{\textsc{IRT-a}} line with a FAP near 10\% (panels B in Fig. \ref{Fig_1}). The C-24 observations have greater time sampling and we find a signal at $P_{\text{a-syn}} = 2.46\,\text{d}$ with a FAP below 1\% in the Ca~{\sc ii}~{\textsc{IRT-a}} line (panel C in Fig. \ref{Fig_1}).

\paragraph{Aliasing}\label{Sec:AliasProblem}

In three of the analyzed epochs, the calcium indicator periodograms show signals compatible with SPI (H-08 at 2.81\,d, C-16 at 2.81\,d, C-24 at 2.46\,d), together with their $\pm$\,1-day aliases. In H-08, the alias is slightly stronger than the primary signal, while in C-16 and C-24, the primary signal dominates. To determine which is the signal and which is the alias, we use the code \texttt{AliasFinder} \cite{Stock_2020}.
This method \cite{DawsonFabrycky2010ApJ...722..937D} has previously been applied to GJ~436~b \cite{Trifon_2018} to determine its radial velocity period (2.64\,d vs. its 1-day alias at 1.6\,d). The weak power of the peaks in all the three epochs analyzed did not yield any conclusive results using \texttt{AliasFinder} (Figs.~\ref{fig:AliasFinder_H08},\ref{fig:AliasFinder_C16},\ref{fig:AliasFinder_C24}). 

\paragraph{Rolling periodograms}\label{Sec:Rolling}
The analysis above was used to identify signals of potential interest, but their statistical significance was low. We use rolling periodograms \cite{Herbort_2018,Schofer_2021} to search for potentially transient periods. Rolling periodograms are calculated from overlapping data subsets. Because the HARPS dataset covers many consecutive years, we first used all the epochs from 2008 to 2020 (Fig. \ref{fig:Rolling_periodogram_HARPS}). For CARMENES, because the two epochs C-16 and C-24 are 8 years apart, we used each of them individually. The rolling periodogram then divides those epochs into smaller sub-epochs of a desired length, computing the GLS periodogram for each of them. If the epoch has $N$ data points and we define each sub-epoch to have $m$ points, and j is the index of the observation's position in the subset, the first sub-epoch is $[j_0, j_m]$. The next sub-epoch is $[j_1, j_{m+1}]$, followed by $[j_2, j_{m+2}]$, and so on, until reaching $[j_{N-m}, j_{N}]$. This allows us to capture variations in the signal across different parts of the split dataset. To select the width of the window m we follow previous work. Previous studies have used 10 datapoints on a 100-day time period \cite{Herbort_2018}, or 21 consecutive datapoints \cite{Schofer_2021}. Another study \cite{Louis_2025} selected a window of 500 days in a time series that spanned from 2019 to 2025, about 22\% of the total timeseries. We adopt that same percentage unless the sub-epoch is shorter than 30 days. We do not allow m $<30$ days, to avoid noise and artifacts. For the HARPS rolling periodogram (Fig.\ref{fig:Rolling_periodogram_HARPS}) we used a 42-day window, while for CARMENES (Fig. \ref{fig:Rolling_periodogram_CARMENES}) we used a 30-day window.\\

For HARPS, out of the 6 epochs that correspond to the 6 years it was observed, two were discarded due to having too few observations ($<10$) to carry out this analysis. From the rolling periodograms of all the combined data, we find that a signal only appears in the epoch corresponding to 2008, at a frequency of $0.356~\text{d}^{-1}$ in Fig.~\ref{fig:Rolling_periodogram_HARPS}. We then used the rolling periodograms in that H-08 single epoch to identify when the SPI signal was strongest, between February 26 and March 11, which is the sub-epoch shown in Fig. \ref{Fig_1}. In the C-16 data (Fig. \ref{fig:Rolling_periodogram_CARMENES}), a signal appears towards the end of the observation window but with less intensity than in the rolling periodograms for HARPS. For the C-24 dataset, due to its shorter time span, we do not find differences between the periodogram of the full dataset or the rolling periodogram.

\paragraph{Statistical significance}\label{subsub:FAP}

In time-series analysis, using the generalized Lomb-Scargle (GLS) periodogram, the  False Alarm Probability (FAP) is used to estimate the likelihood that a detected signal could arise purely from random noise. In a GLS periodogram, lower FAP levels indicate higher confidence in the detection of a peak: for example, a FAP of 1\% indicates a 1\% chance that random noise alone could have produced a signal of that strength at any frequency. However, global FAP calculations assume that we are searching blindly over many frequencies without prior knowledge of where a signal might occur, which makes them less appropriate when the frequency of interest is already known from physical considerations. In our analysis, we instead consider how likely it is that random noise could produce a peak of that strength at that specific frequency.\\

The global FAP (as shown in Fig. \ref{Fig_1}) does not account for this distinction, nor for possible correlated noise (red noise) and irregular sampling. We adopt bootstrap randomization in conjunction with a windowing technique \cite{Hatzes_2019}. The bootstrapping randomly shuffles the observed data points many times to generate synthetic datasets that preserve the original noise properties and sampling times. We then test how often a peak as strong as the observed one appears at the expected frequency. To determine the local FAP at a specific frequency, we repeated this process $5\times10^{6}$ times while performing the bootstrap in frequency windows centered on the frequency of interest.  We start with a large frequency window of 3 times the full width at half maximum (FWHM) of the peak, then repeat the analysis with smaller widths (Fig. \ref{fig:FAP_extrapolation}). We fitted the points with a low order polynomial, which was extrapolated to zero window width to determine the FAP.
This bootstrap-based FAP provides an empirical estimate of the probability that random noise could mimic a true signal at an expected frequency. The resulting local FAPs for each expected period in each epoch are listed in Table \ref{FAP-table}.

The per-epoch FAPs in Table \ref{FAP-table} quantify the local significance of each individual detection at its expected frequency, but they do not address the overall significance of the observed pattern of detections across the full search. The statistically appropriate global significance must account for the fact that we analyzed seven spectroscopic epochs in total (H-07, H-08, H-09, H-10, H-20, C-16, C-24), including four in which no significant peak was found, as well as for the geometric model's prior that signals can fall at either $P_{\rm syn}$ or $P_{\rm a-syn}$. We obtain this trials-corrected global significance by applying Fisher's combined probability test \cite{Fisher_1925}, the standard procedure for combining independent p-values into a single global probability. Under the null hypothesis of no SPI signal, each properly computed p-value is uniformly distributed on $[0,1]$, so the transformed quantity $-2\ln p$ follows a $\chi^2$ distribution with two degrees of freedom, the sum over $k$ independent tests is then distributed as $\chi^2_{2k}$. The logarithmic transformation weights small p-values heavily, while non-detections contribute small positive terms ($-2\ln 0.5 \approx 1.4$), so the four non-detection epochs appropriately dilute the global statistic rather than being ignored. The independence assumption is justified by the temporal separation between epochs, five seasons spread over more than a decade for HARPS, and two separated by eight years for CARMENES, and by the bootstrap procedure, which independently randomizes any time-correlated structure within each epoch.

We applied this test uniformly to all seven epochs. For each epoch we computed the maximum GLS periodogram power in the period range $[2.36, 3.0]$~d, which brackets both frequencies predicted by the geometric model ($P_{\rm syn} = 2.81$~d and $P_{\rm a-syn} = 2.46$~d). The per-epoch p-value was obtained by bootstrap shuffling of the (value, error) pairs, preserving the original timestamps and noise properties ($10^5$ shuffles per epoch). We do not penalise the calculation for trials over activity indicators, since our focus on Ca\,\textsc{ii} is motivated a priori by the formation-region argument given in the main text.

The resulting per-epoch p-values and their contributions to the Fisher statistic are summarised in Table \ref{table:fisher}. Figure \ref{fig:Fischer_per_epoch} shows the bootstrap null distribution of the maximum power in $[2.36, 3.0]$~d for each epoch alongside the observed value, demonstrating that the three detection epochs (H-08, C-16, C-24) sit in the upper tails of their respective null distributions, while the four non-detection epochs sit well within the bulk. The combined Fisher statistic is
\begin{equation}
X^2_{\rm obs} = -2 \sum_{i=1}^{7} \ln p_i = 33.78,
\end{equation}
with $2N_{\rm epoch} = 14$ degrees of freedom. The corresponding global p-value, computed analytically from the $\chi^2_{14}$ survival function, is $p_{\rm global} = 2.22 \times 10^{-3}$. The observed $X^2_{\rm obs}$ exceeds the 99.7\% critical value of $\chi^2_{14}$ ($X^2_{\rm crit} = 32.88$).

To verify this analytical result, we also computed the global p-value by direct Monte Carlo resampling. For each of $10^6$ synthetic trials, we drew one random max-power value per epoch from its own bootstrap null distribution, converted these draws to p-values using the same rank procedure as the observed data, and computed the resulting $X^2$. The empirical fraction of synthetic trials exceeding $X^2_{\rm obs}$ is $p_{\rm global} = 2.18 \times 10^{-3}$, in agreement with the analytical result to within Monte Carlo precision. The full empirical null distribution and its analytical $\chi^2_{14}$ prediction are shown in Fig. \ref{fig:Fischer}, the agreement between the two confirms that the per-epoch bootstrap p-values are well-calibrated and that the assumptions underlying Fisher's test are met.

\paragraph{Other periodograms methods}
\paragraph{$\ell$-1 periodogram}
The $\ell$-1-norm-based periodogram, hereafter the $\ell$-1 periodogram \cite{Hara_2016} searches for periodic signals in unevenly sampled time series, similar to the Lomb-Scargle \cite{Scargle_1982} periodogram, but less prone to aliasing. We use this periodogram to analyze the epochs and find identical results to those obtained from the GLS.

\paragraph{PDC, \texttt{USuRPER} and \texttt{SPARTA} periodograms}
The Phase Distance periodogram (PDC) \cite{Shay_2017} is a model-independent tool for periodicity detection in astronomical time series with unevenly sampled data. \texttt{USuRPER}, is an extended version of PDC that allows the period to vary over time \cite{Binnenfeld_2020}.\\

We used the Spectroscopic Variability Analysis (\texttt{SPARTA}) \cite{Shahaf_2020}, implementation of \texttt{USuRPER} to apply these periodograms to all the epochs, including H-08, C-16 and C-24 where we found no signal at the $P_{\text{syn}}$ or $P_{\text{a-syn}}$ periods. The results from this analysis showed periodic signals at 2.24 and 1.79\,d  corresponding to non-Doppler spectral variations.\\

\subsubsection*{Modelling of the signal} \label{Modeling-sec}

We propose a geometric model to explain an SPI signal at two distinct periods that differ from the planetary orbital period. We hypothesize that the intensity of the signal changes with the distance between two active regions in the chromosphere of the star. One of these active regions has a stellar origin and thus rotates with the star (point A), while the other is induced by SPI and moves according to the planet’s orbit (point P). As a result of the different rotation period, both regions are not always visible to an observer. The region with shorter period (we assume point P), moves faster on the chromosphere. Combined observations of those two separated and independent regions would cause $P_{\text{orb}}$ to split into $P_{\text{syn}}$ and $P_{\text{a-syn}}$.\\

We consider a reference frame with its origin $O$ at the center of the star and the $XY$ plane coincident with the equatorial plane of the star. The orbit of GJ~436~b is highly inclined and moderately eccentric. For simplicity, we consider only the effect of the obliquity of the planetary orbit $I$, the angle between the equatorial plane of the star and the plane of the orbit, which corresponds to  $I = 103^{\circ} \pm 13^{\circ}$ for GJ~436~b \cite{Bourrier_2018}. This is the 3D obliquity, not the obliquity projected on the plane of the sky. We assume that the $X$ axis is directed along the line of nodes of the planetary orbit, pointing towards the ascending node. Considering the coordinate transformation (Fig.~\ref{fig:Coordinate}), and assuming the orbit to be circular with an orbital radius $a$, the coordinates of the planet $(X_{\rm p}, Y_{\rm p}, Z_{\rm p})$ are

\begin{equation}
    \left\{ 
\begin{array}{lll}
X_p &= a \cos (f+\phi_1), \\
Y_p &= a \sin (f+\phi_1) \cos I, \\
Z_p &= a \sin (f+\phi_1) \sin I
\end{array}
\right. 
\end{equation}
where $f = n t$ is the orbital phase of the planet along its orbit with $n \equiv 2\pi/P_{\rm orb}$ being its mean angular frequency, $P_{\rm orb}$ the orbital period, and $t$ the time measured from the passage through the ascending node.\\

Assuming the active region $A(X,Y,Z)$ is located on the equator of the star

\begin{equation}
    \left\{ 
\begin{array}{lll}
X &= R \cos (\lambda +\phi_2), \\
Y &= R \sin (\lambda +\phi_2), \\
Z & = 0
\end{array}
\right. 
\end{equation}
where $R$ is  the radius of the star, $\lambda = \lambda_{0} + \Omega t$ with $\lambda_{0}$ the longitude of the point $A$ at time $t=0$, $\Omega = 2\pi/P_{\rm rot}$ is the stellar spin angular velocity, and $P_{\rm rot}$ is the stellar rotation period. One of the two phases ($\phi_1$ or $\phi_2$) can be set to zero by an appropriate choice of the origin of the coordinate system; in this case, we set $\phi_2 = 0$.\\

We hypothesize that the SPI signal depends on the distance $d = {AP}$. Substituting the coordinates of the points $A$ and $P$ into the distance formula, we find:

\begin{equation}
    \begin{array}{l}
 {AP}^{2} = (X-X_{\rm p})^2 + (Y-Y_{\rm p})^{2} + Z_{p}^{2} = \\
=  R^{2} + a^{2} -2aR \left( \cos \lambda \cos (f+\phi_1) + \sin \lambda \sin (f+\phi_1) \cos I \right) = \\
= R^{2} + a^{2} - 2aR \cos (\lambda-f-\phi_1) +2 aR (1 - \cos I) \sin \lambda \sin (f+\phi_1).
    \end{array}
\end{equation}\\

The difference $\lambda - f = (\Omega -n) t + \lambda_{0}$, while the product $\sin \lambda \sin f =  \left[ \cos(\lambda-f) - \cos(\lambda+ f) \right]/2$ by applying Werner's formulae. Then:

\begin{equation}
\begin{array}{l}
    {AP}^{2} = R^{2} + a^{2} - 2 aR \cos \left[ (\Omega-n) t + \lambda_{0 }-\phi_1 \right] + \\
    aR (1 -\cos I) \left\{ \cos \left[ (\Omega-n)t + \lambda_{0}-\phi_1 \right] - \cos \left[ (\Omega+n) t +\lambda_{0}+\phi_1 \right] \right\}. 
\end{array}
    \label{dist2}
\end{equation}\\

If $R=a$ this expression would describe the motion of a point on the star that is following the motion of the planet along its orbit, the sub-planetary point. Then, Eq.~\ref{dist2} shows that the distance ${AP}$ oscillates at the synodic frequency $\Omega -n$ when the obliquity of the orbit is zero, because $I = 0$ implies $\cos I =1$. On the other hand, for an oblique orbit as in the case of GJ~436~b, $I \not= 0$, and the other frequency $\Omega + n$ appears. Therefore, the appearance of the frequency $\Omega +n$ in the SPI periodicity of GJ~436 could be related to the obliquity of its orbit. This geometrical model provides a potential explanation of why that frequency was not previously observed in other systems: because those orbits were aligned with the equatorial plane of the host star.\\ 

The orbit of GJ~436~b has eccentricity $e = 0.152$ \cite{Vidotto_2023}. Introducing the eccentricity in the above model complicates the mathematical development because both the orbital radius (the distance between $O$ and $P$) and the longitude of the planet along its orbit $f$ become complex functions of time. We therefore regard our assumption of a circular orbit as a first-order approximation.

\paragraph{Alternating occurrence of the signals}

To explore the behavior of Eq.~\ref{dist2}, we generate synthetic datasets with the same number of points as our observations but with the timestamps distributed randomly in the same range of dates as the original observations. Uncertainties were generated from a separate Gaussian distribution with the same standard deviation as the C-24 data.\\

By randomly generating the timestamps, we observe that different distributions of the data points produce periodograms with either $P_{\text{syn}}$ or $P_{\text{a-syn}}$ (Fig.~\ref{Fig_2}). Regardless of the timing of the observations, the orbital period $P_{\rm orb}=2.64$\,d does not appear in the periodogram. The other period of interest, $P_{\rm rot}$, could appear in the interaction's resulting periodogram as seen in our synthetic data (Fig.\ref{Fig_2}). We do not find this signal in the calcium indicators, but it is present in the H$\alpha$ line (Fig. \ref{CARMENES_2024_All}).\\

The change in the detected period between epochs might be related to the sampling of the data. In data sets with fewer data points, one of the frequencies can dominate the periodogram due to windowing effects or incomplete coverage of the phase space. As the number of data points increases and the sampling becomes denser and more uniform, both frequencies could be recovered and appear clearly resolved in the periodogram.

\subsubsection*{Stellar properties}

\paragraph{Activity cycle} \label{Activity_cycle}
Variations in the host star magnetic field strength, stellar wind, and the surrounding stellar medium during its stellar activity cycle could influence the detection and characterization of SPI.\\

GJ 436's activity cycle has previously been determined using photometry from T12 APT, which showed a modulation of 7.4 years \cite{Lothringer_2018}. Other studies using the same photometric data determined a slightly longer period of $7.75 \pm 0.10$ years \cite{Loyd_2023}. HARPS spectroscopic observations have shown a period of approximately 6.8 years \cite{Kumar_2022}. 
We reconsider these values because an additional T12 APT epoch (2017-18) is not consistent with them. This data shows an increase in GJ 436's brightness which could imply an increase in the activity and spot coverage of the star. We use observations in the  Cousins $R$ passband using C14 AIT. Even though the T12 APT and C14 AIT observe in different bands, we combine both datasets by adding an offset to the AIT data, calculated from the mean values of the epochs in which both facilities were observing. The combined light curve (Fig.~\ref{Fig_3} B) shows a linear trend toward higher brightness. The amplitude of the modulation  seen in T12 APT data is not apparent in the C14 AIT data taken after 2020. Since we have contemporary monitoring of the star with both instruments (C14 AIT and T12 APT) during several epochs ($\sim$ 2014 to $\sim$ 2017), we ascribe this change to the star rather than the instrument.\\

We compare the photometric and chromospheric long-term time series and found an anti-correlation between photometric variability and chromospheric activity (Fig.~\ref{Fig_3}), in line with previous studies \cite{dos_Santos_2019,Loyd_2023}. This anti-correlation indicates a spot-dominated stellar magnetic activity. In spot-dominated stars, photometric brightness is at its minimum during the activity maximum, because a larger portion of the stellar surface is covered by spots, which emit less light in the wavelengths typically observed in photometry. Conversely, activity indicators, such as the Ca~{\sc ii}~{\textsc{H\&K}} lines, reach their maximum, consistent with our observations of GJ 436 (Fig.~\ref{Fig_3}).

\paragraph{Rotation period} \label{Sec:Rotation}
Photometric observations with T12 APT over 14 years revealed a rotation period of $P_{\rm rot} = 44.09 \pm 0.08$\,d \cite{Bourrier_2018}. A variability in the period was observed across different epochs, ranging between 41.7\,d to 46.6\,d \cite{Bourrier_2018}. Subsequent independent analyses of these observations confirmed this rotation period \cite{Lothringer_2018, Loyd_2023}. Other photometric observations \cite{Pollaco_2006} showed a rotation period of $P_{\rm rot} = 44.6 \pm 2$\,d \cite{DiezAlonso_2019}. Spectroscopic activity indicators such as $H\alpha$ and Ca~{\sc ii}~{\textsc{H\&K}} are consistent with the photometric rotation periods, with values ranging from 40 to 50 days \cite{Bourrier_2018,Lothringer_2018,Kumar_2022}. \\

\begin{figure}
\centering
\includegraphics[width=\columnwidth]{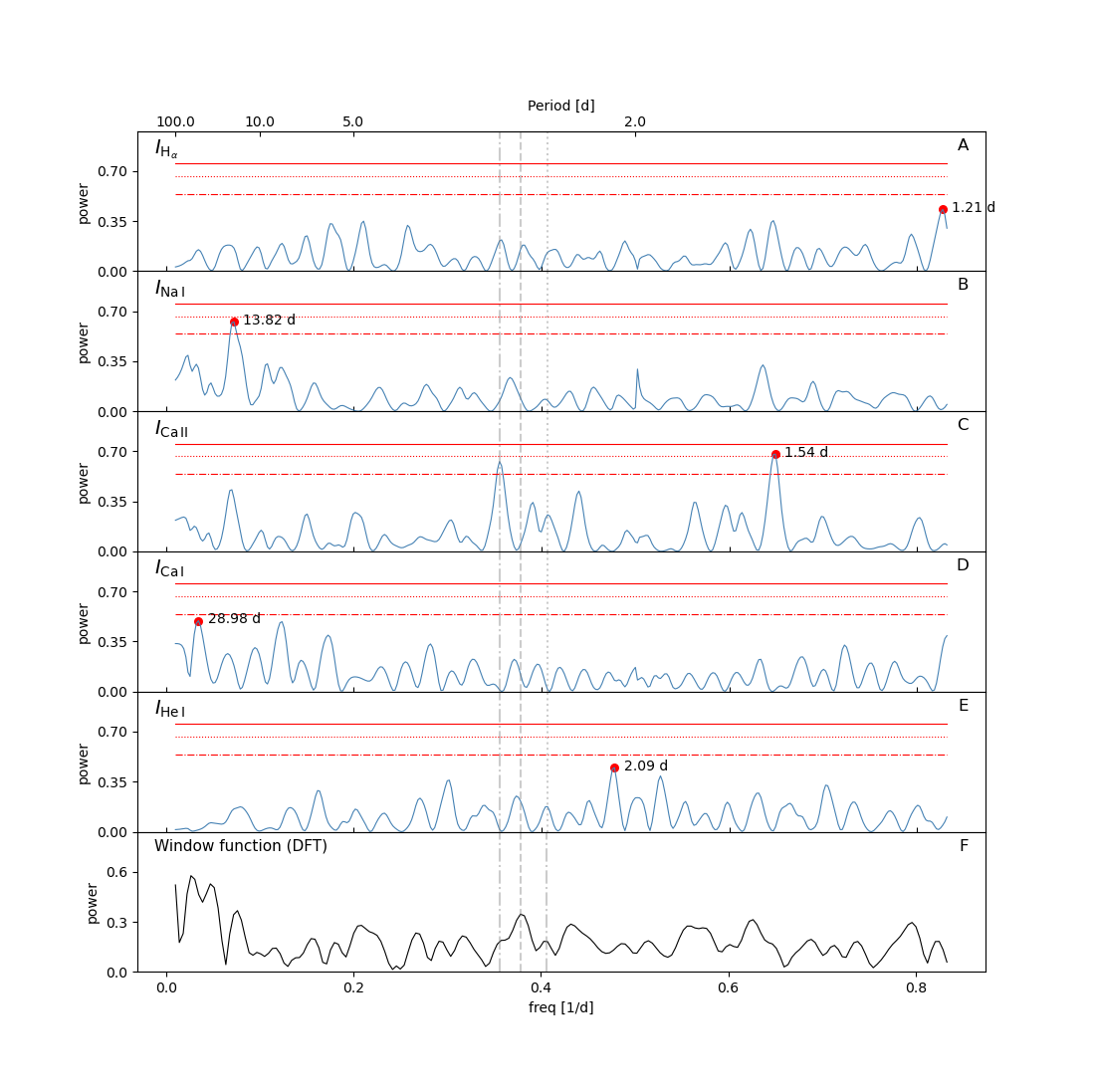}
\caption{\textbf{Same as Fig. \ref{HARPS_2008full_All} but for H-08 sub-epoch activity indicators measured using HARPS.}}
\label{HARPS_2008_All}
\end{figure}

\begin{figure}
    \centering
    \includegraphics[width=\columnwidth]{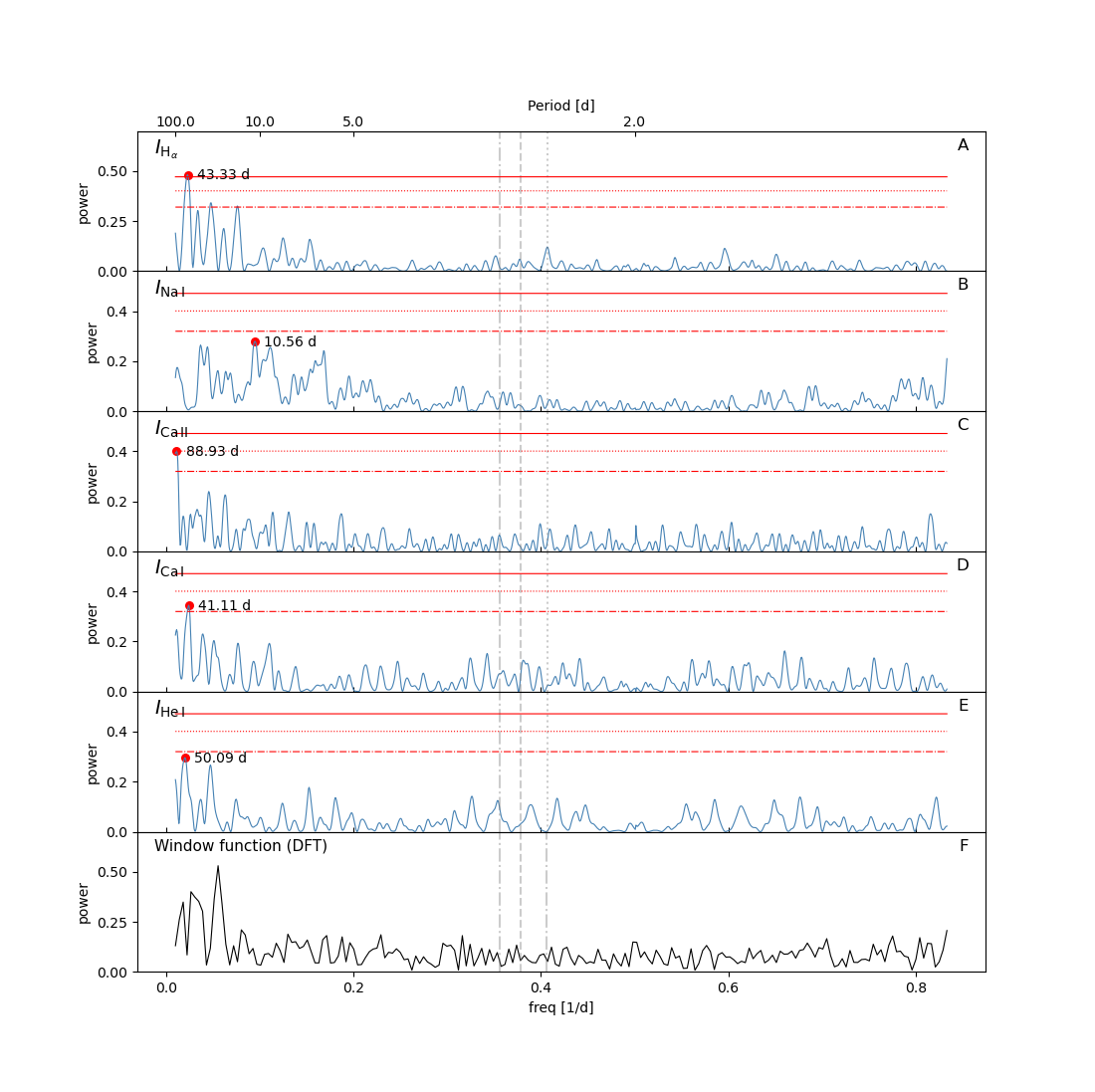}
    \caption{\textbf{Same as Fig.~\ref{HARPS_2007full_All}, but covering the entire 2009 observational season.}}
    \label{HARPS_2009_All}
\end{figure}

\begin{figure}
    \centering
    \includegraphics[width=\columnwidth]{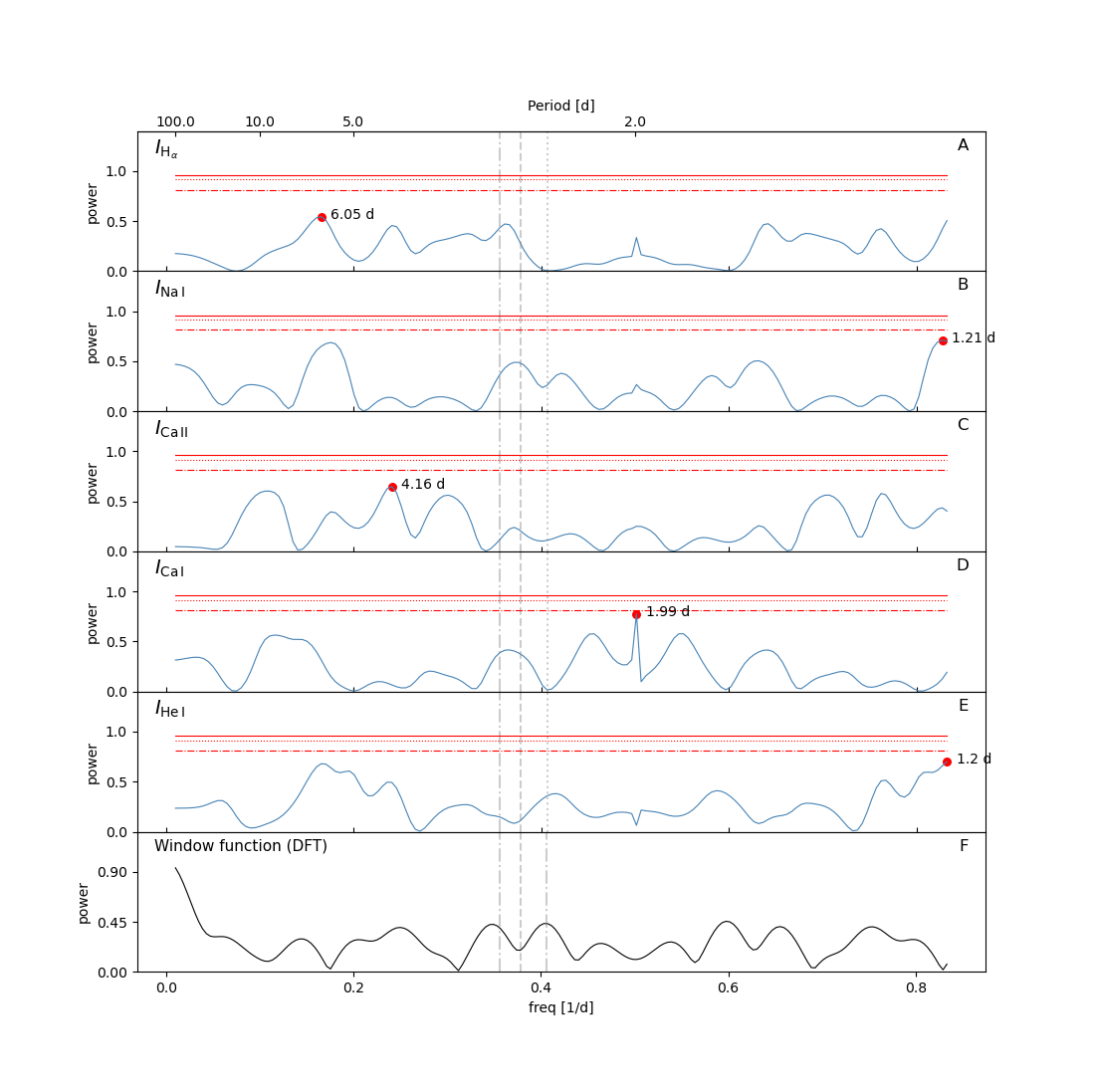}
    \caption{\textbf{Same as Fig.~\ref{HARPS_2007full_All}, but covering the entire 2010 observational season.}}
    \label{HARPS_2010_All}
\end{figure}

\begin{figure}
    \centering
    \includegraphics[width=\columnwidth]{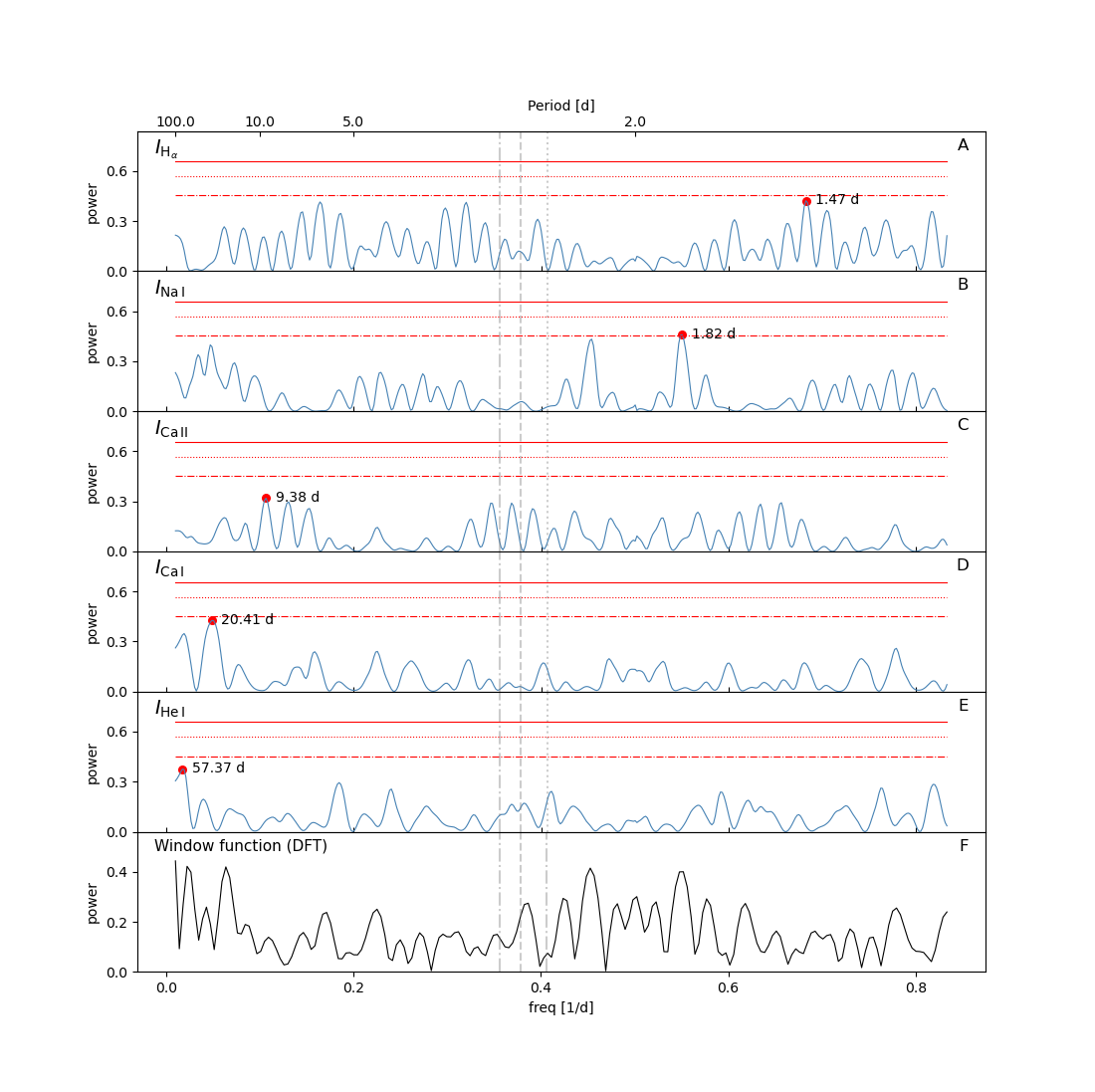}
    \caption{\textbf{Same as Fig.~\ref{HARPS_2007full_All}, but covering the entire 2020 observational season.}}
    \label{HARPS_2020_All}
\end{figure}

\begin{figure}
\centering
\includegraphics[width=\columnwidth]{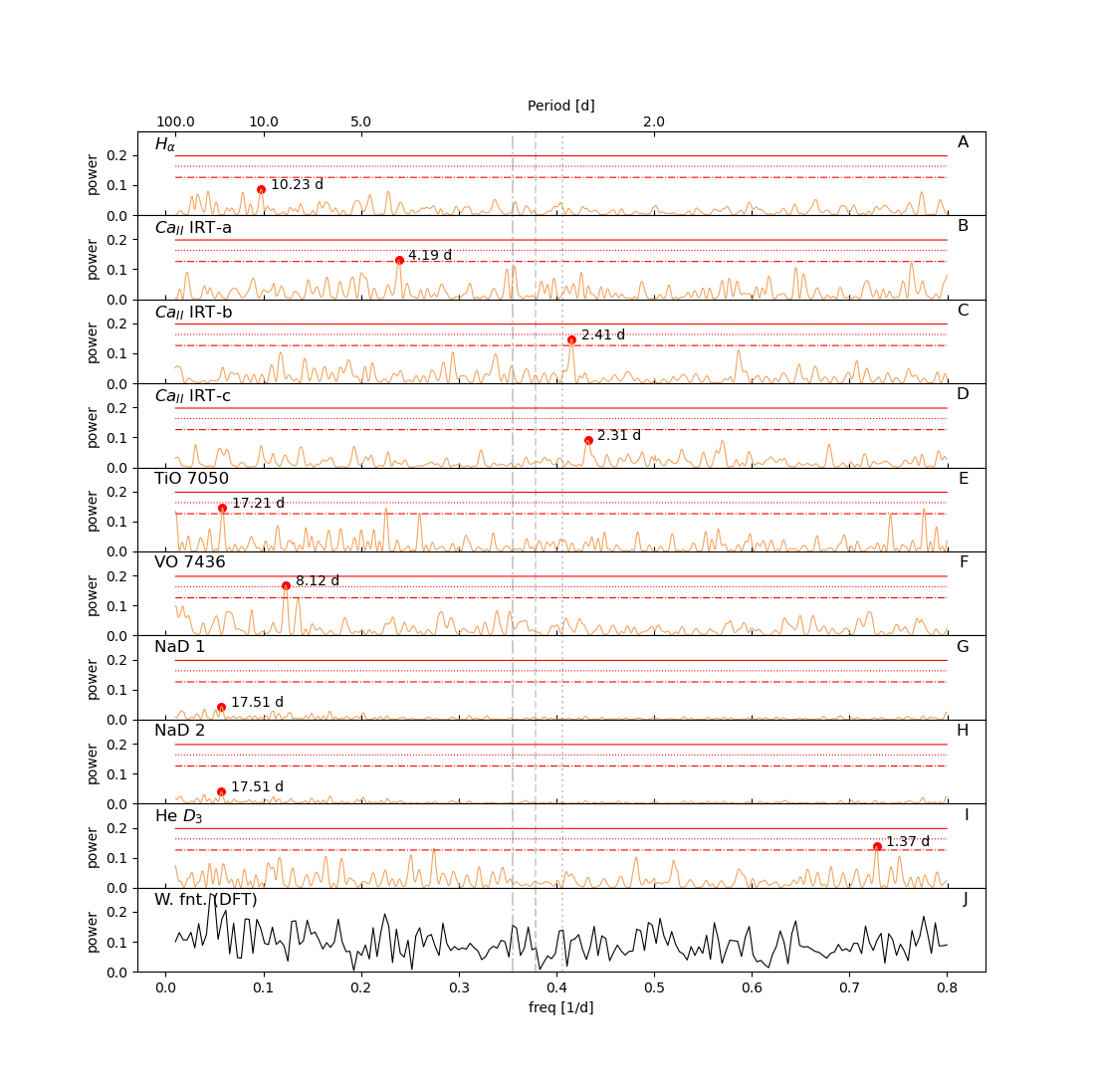}
\caption{\textbf{GLS periodogram for C-16 sub-epoch, including all activity indicators measured with CARMENES.} The periodogram is plotted in orange following the same color scheme as in Fig. \ref{Fig_1}, where CARMENES periodograms are plotted in orange while HARPS periodograms in blue.}
\label{CARMENES_2016_All}
\end{figure}

\begin{figure}
\centering
\includegraphics[width=\columnwidth]{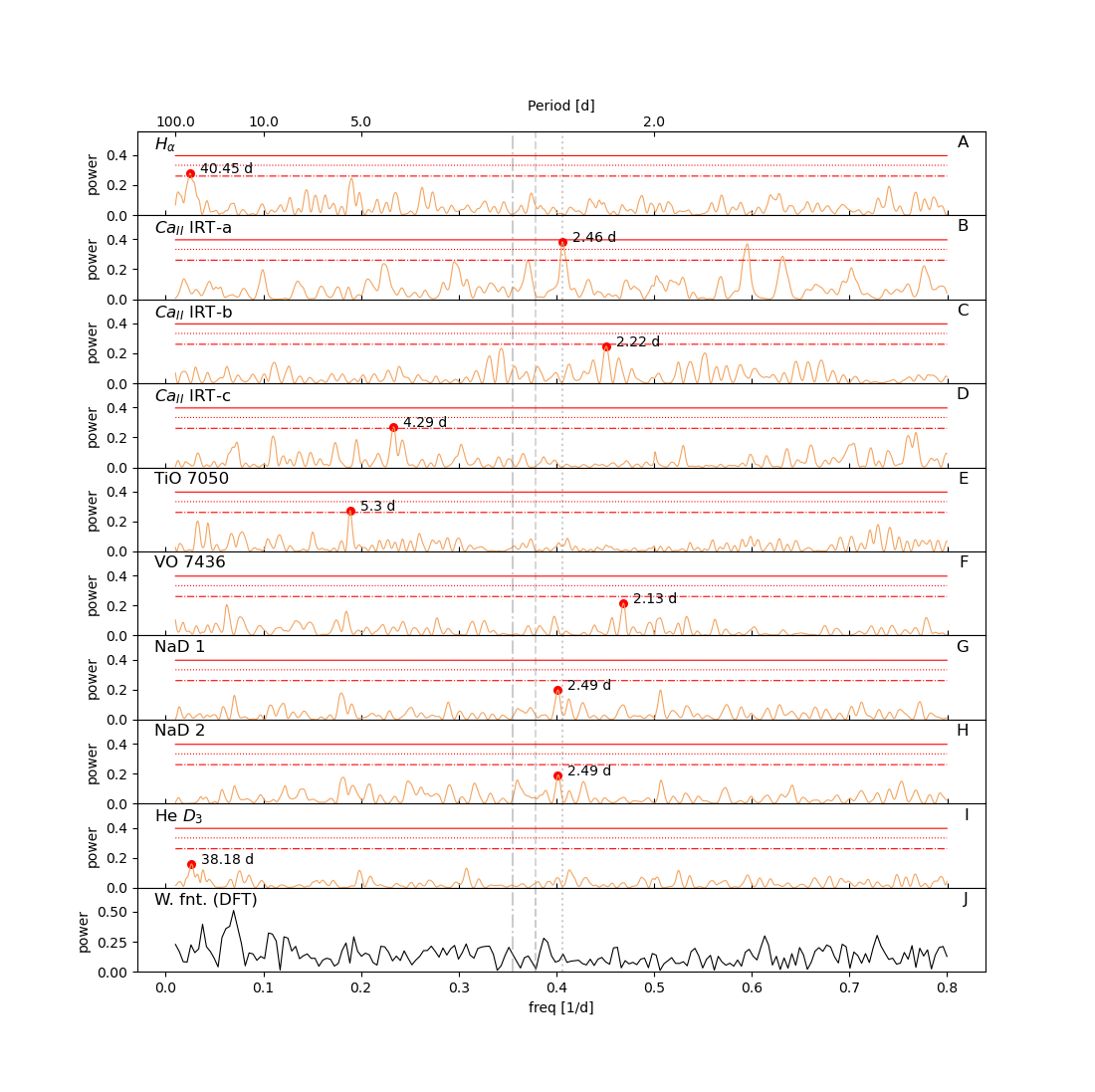}
\caption{\textbf{Same as Fig.~\ref{CARMENES_2016_All}, but for C-24.}}
\label{CARMENES_2024_All}
\end{figure}

\begin{figure}
    \centering
    \includegraphics[width=1\linewidth]{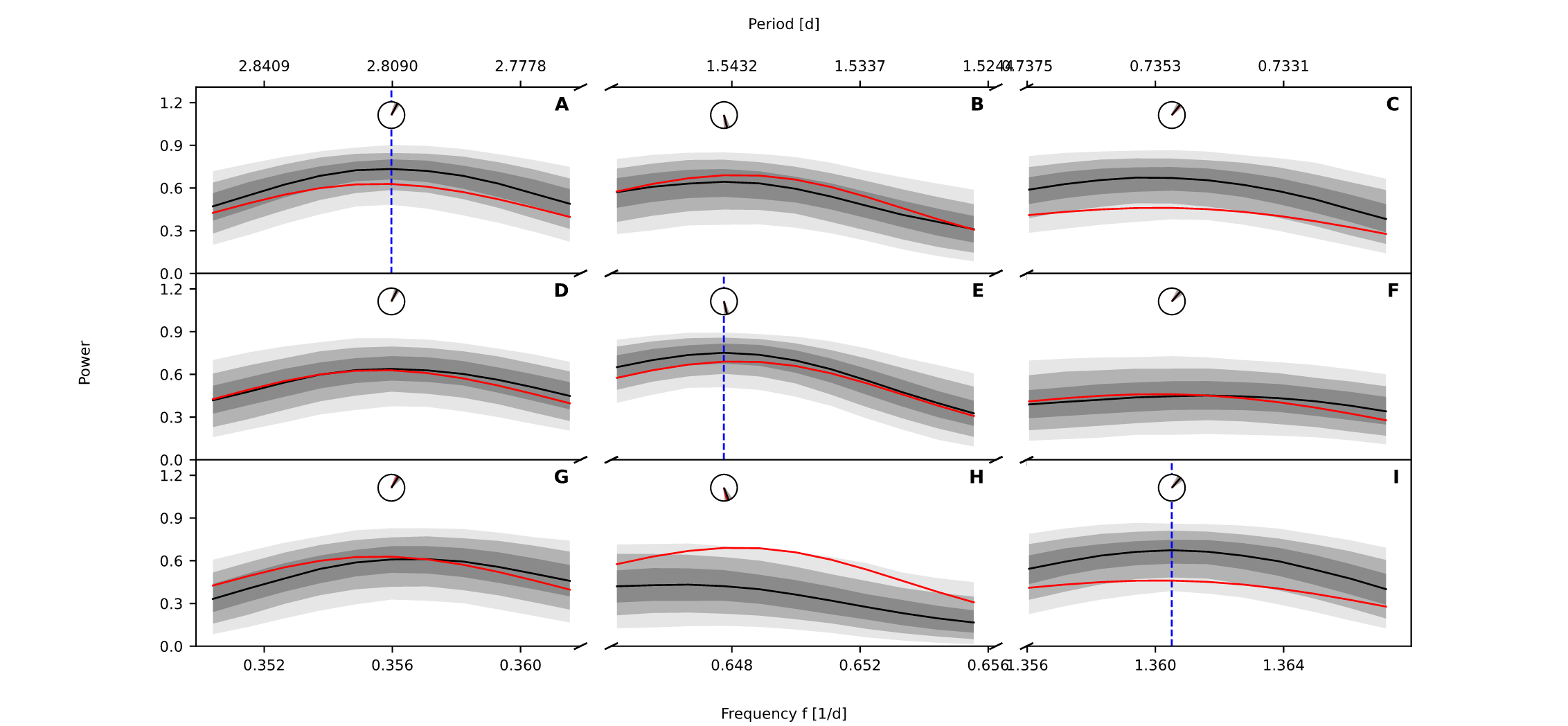}
    \caption{\textbf{Aliasing analysis with \texttt{AliasFinder}.} (\textbf{A}): The periodogram of the observed data (red) compared to the median GLS periodogram (dark) from simulations with an injected signal at 2.81\,d  (vertical blue dashed line). The darkest shaded grey areas show the interquartile range (25th–75th percentile), the medium the 80\% range (10th–90th percentile), and the lightes the 95\% range (2.5th–97.5th percentile) of the simulated periodograms. The clocks show the angular mean of the phase of each peak and its standard deviation. (\textbf{B},\textbf{C}) The regions around the 1-day aliases of the period in panel A: 1.55\,d and 0.73\,d. The other rows are the same as panels A-C but injecting a signal at (\textbf{D-F}) 1.55\,d and  (\textbf{G-I}) 0.73\,d, respectively. }
    \label{fig:AliasFinder_H08}
\end{figure}

\begin{figure}
    \centering
    \includegraphics[width=1\linewidth]{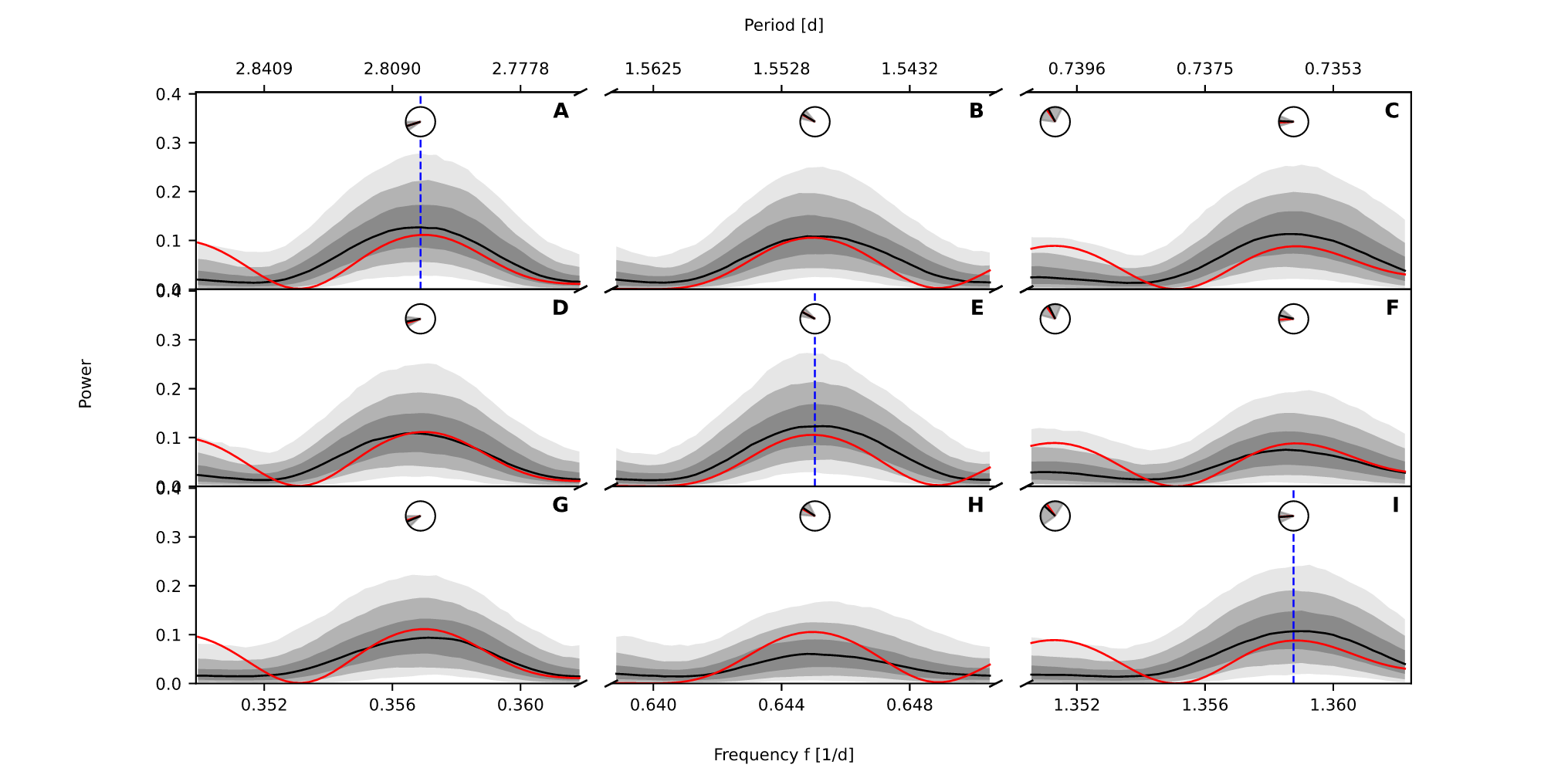}
    \caption{\textbf{Same as Fig. \ref{fig:AliasFinder_H08} but for the C-16 dataset.}}
    \label{fig:AliasFinder_C16}
\end{figure}

\begin{figure}
    \centering
    \includegraphics[width=1\linewidth]{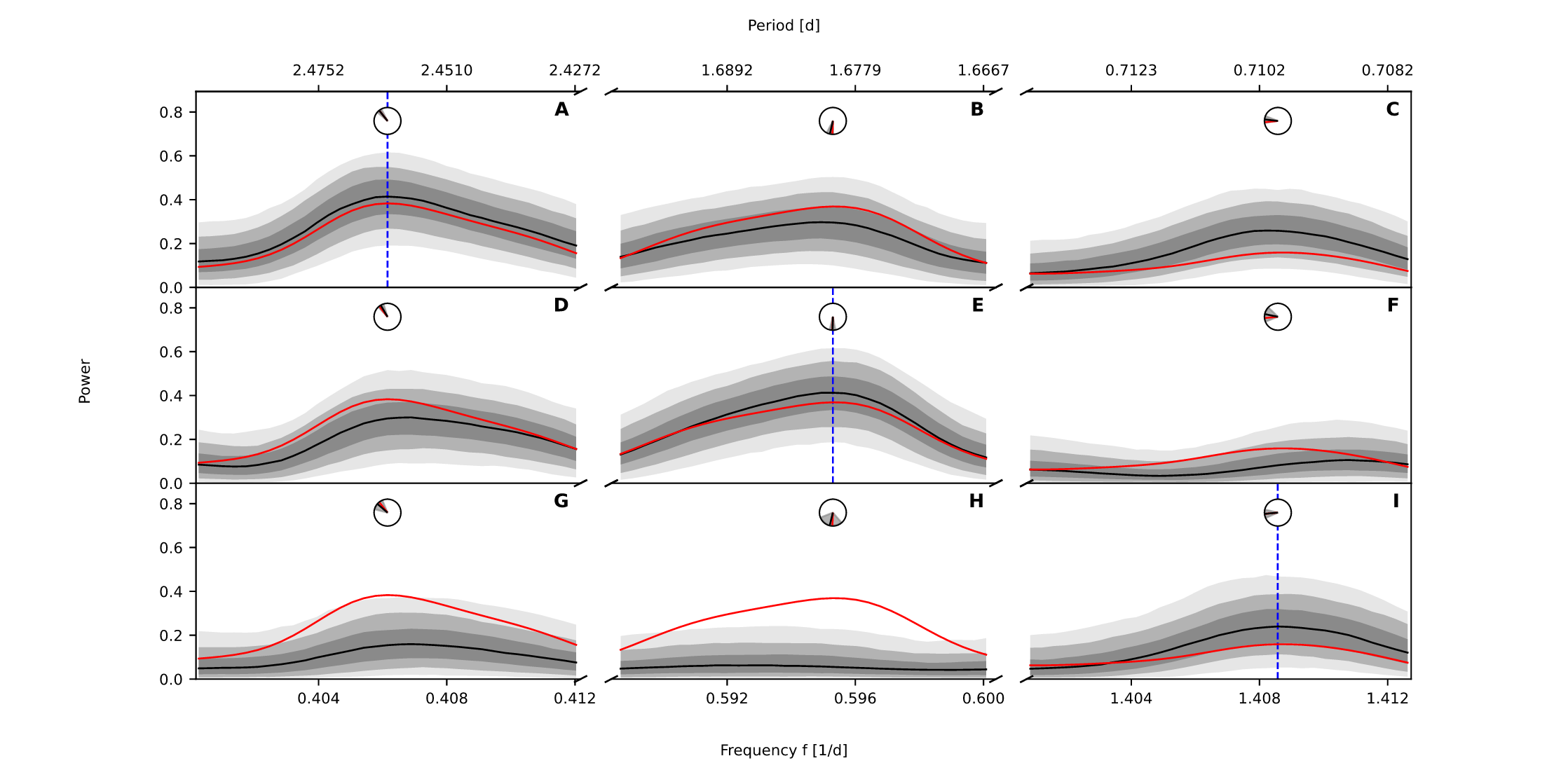}
    \caption{\textbf{Same as Fig. \ref{fig:AliasFinder_C16} but for the C-24 dataset, injecting a signal of 2.46\,d,  1.68\,d and 0.71\,d }}
    \label{fig:AliasFinder_C24}
\end{figure}

\begin{figure}
\centering
\includegraphics[width=0.9\columnwidth]{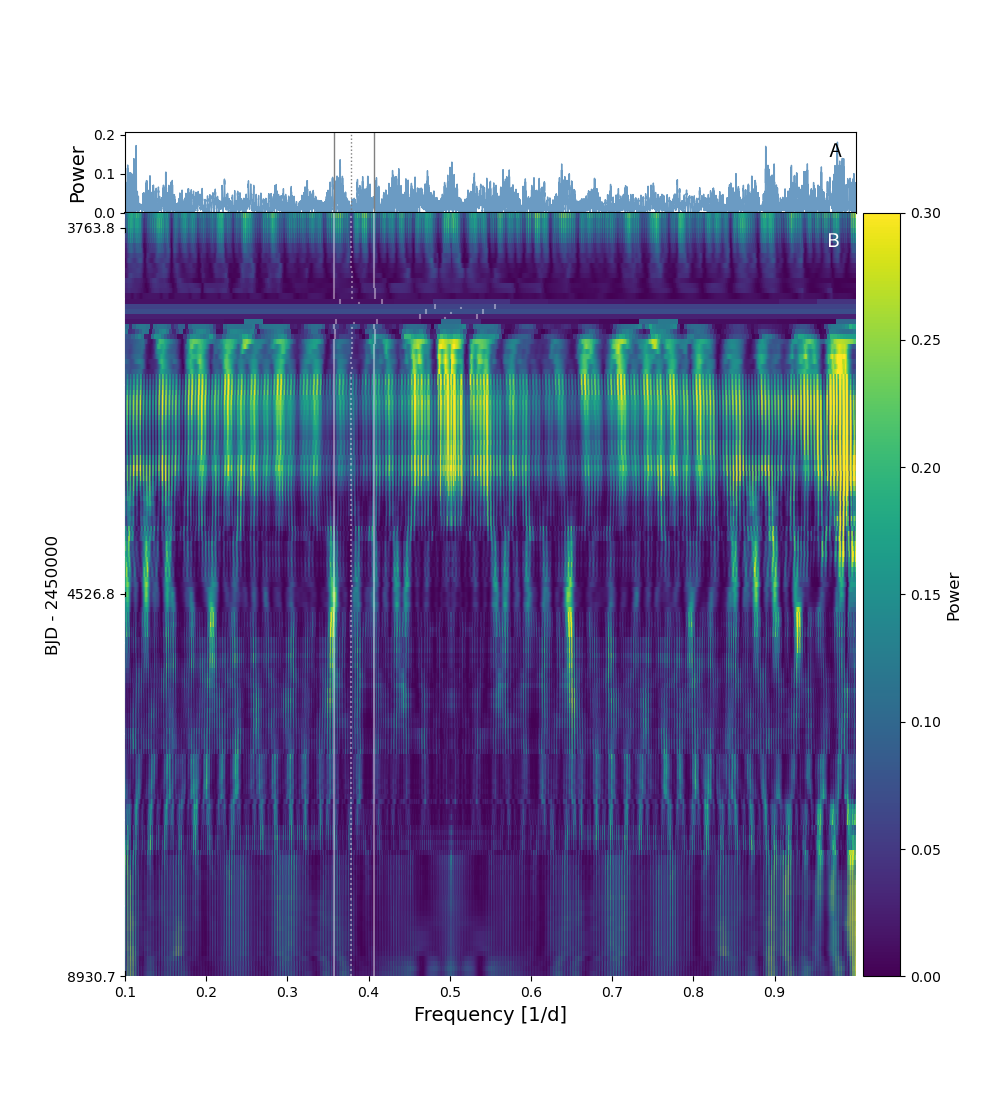}
\caption{\textbf{Rolling periodogram of $I_{\rm CaII}$ from all HARPS observations taken between 2006 and 2020.} \textbf{(A):} In blue, the GLS periodogram of the entire HARPS dataset. Three vertical lines indicate the $P_{\text{syn}}$ (left solid), $P_{\text{orb}}$ (dotted) and $P_{\text{a-syn}}$ (right solid). \textbf{(B):}  Rolling periodogram of $I_{\rm CaII}$ computed over a sliding window of $\text{m}=42$ consecutive observations stepped through the full baseline. The color scale shows the GLS power, with brighter (yellow) regions indicating higher significance. A persistent, coherent signal is visible near $P_{\rm syn} = 2.81$ d throughout the middle portion of the time baseline (2008 epoch). Three vertical lines indicate the $P_{\text{syn}}$ (left solid), $P_{\text{orb}}$ (dotted) and $P_{\text{a-syn}}$ (right solid).}
\label{fig:Rolling_periodogram_HARPS}
\end{figure}

\begin{figure}
\centering
\includegraphics[width=\columnwidth]{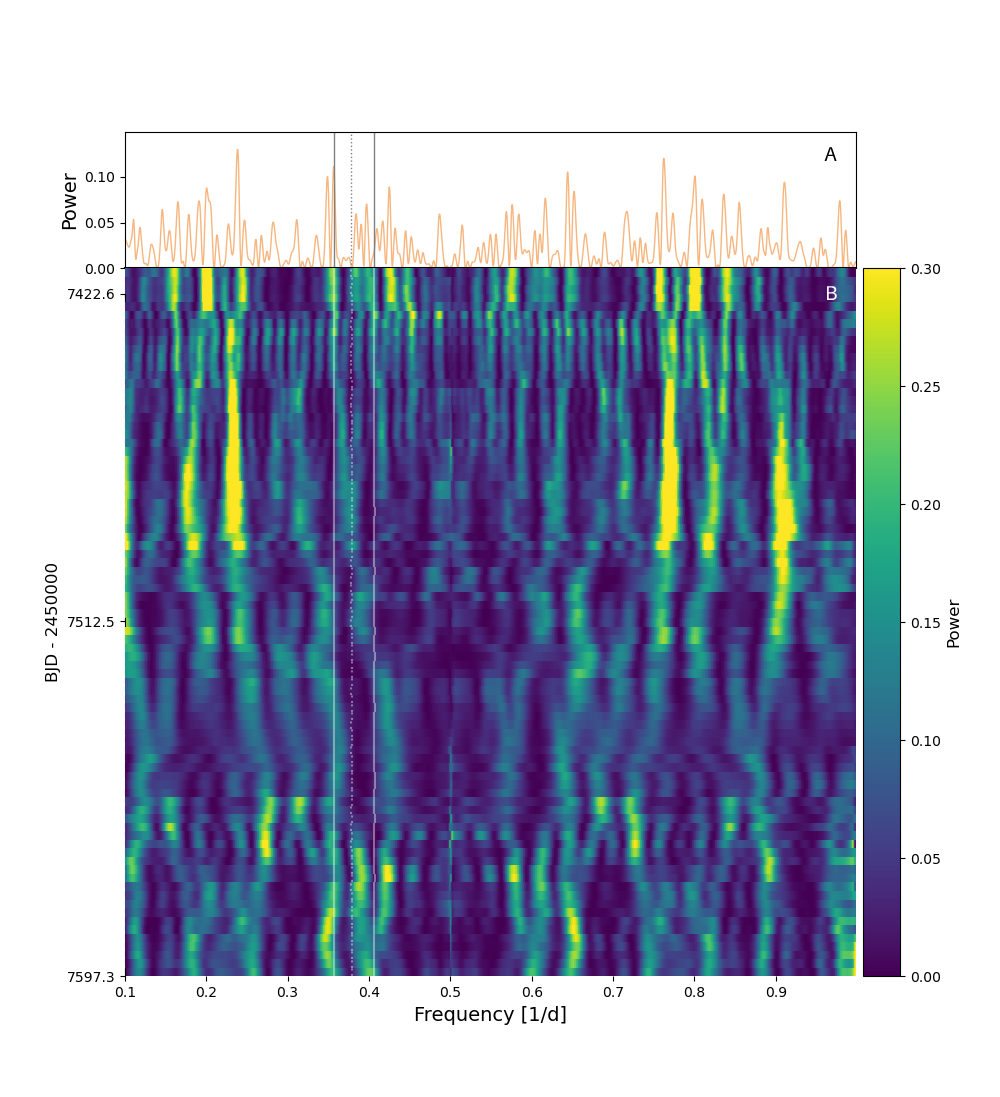}
\caption{\textbf{Same as Fig.~\ref{fig:Rolling_periodogram_HARPS}, but for the pEW'(Ca~{\sc ii}~{\textsc{IRT-a}}) in the 2016 epoch observed with CARMENES.}}
\label{fig:Rolling_periodogram_CARMENES}
\end{figure}

\begin{figure}
    \centering
    \includegraphics[width=0.8\linewidth]{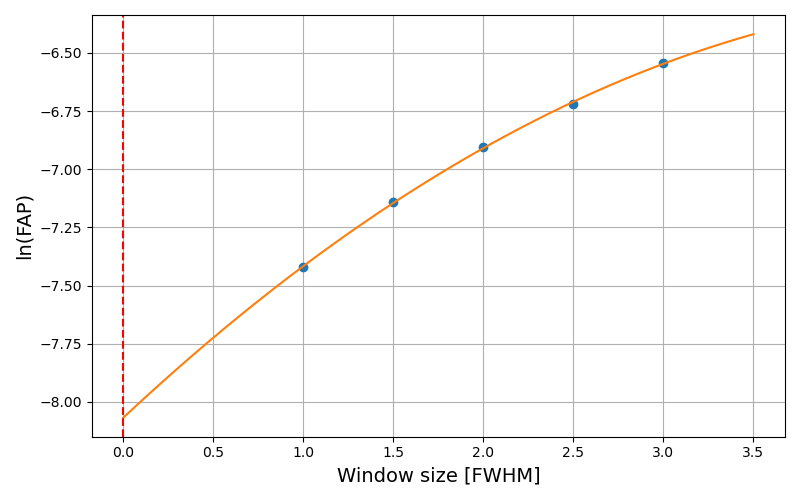}
    \caption{\textbf{Windowing bootstrap technique to extrapolate the local FAP at a specific frequency.} The blue dots are the ln(FAP) obtained for a specific window size, in multiples of the FWHM. The orange line is a quadratic model fitted to the data points. Its extrapolation to x=0 (red dashed line) is the local FAP of a specific expected frequency.}
    \label{fig:FAP_extrapolation}
\end{figure}

\begin{figure}
    \centering
    \includegraphics[width=1\linewidth]{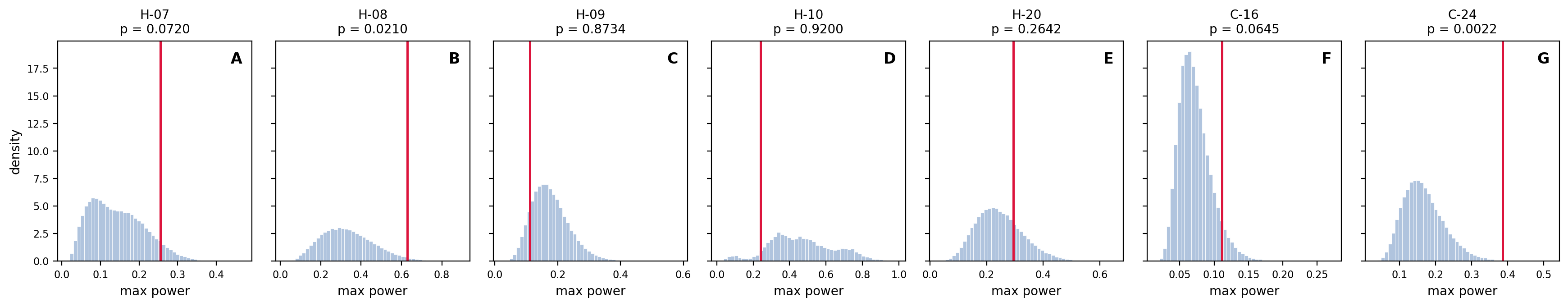}
    \caption{\textbf{Per-epoch bootstrap null distributions.} 
(\textbf{A--G}): The blue histogram shows the null distribution of the maximum periodogram power in the search range for each of the seven spectroscopic epochs. The red vertical line marks the maximum power observed in the data. The resulting per-epoch p-value is given above each panel. The three detection epochs (\textbf{B}: H-08, \textbf{F}: C-16, \textbf{G}: C-24) place the observed value in the upper tail of their respective null distributions, while the four non-detection epochs (\textbf{A}: H-07, \textbf{C}: H-09, \textbf{D}: H-10, 
\textbf{E}: H-20) place the observation within the bulk.}
    \label{fig:Fischer_per_epoch}
\end{figure}

\begin{figure}
    \centering
    \includegraphics[width=0.8\linewidth]{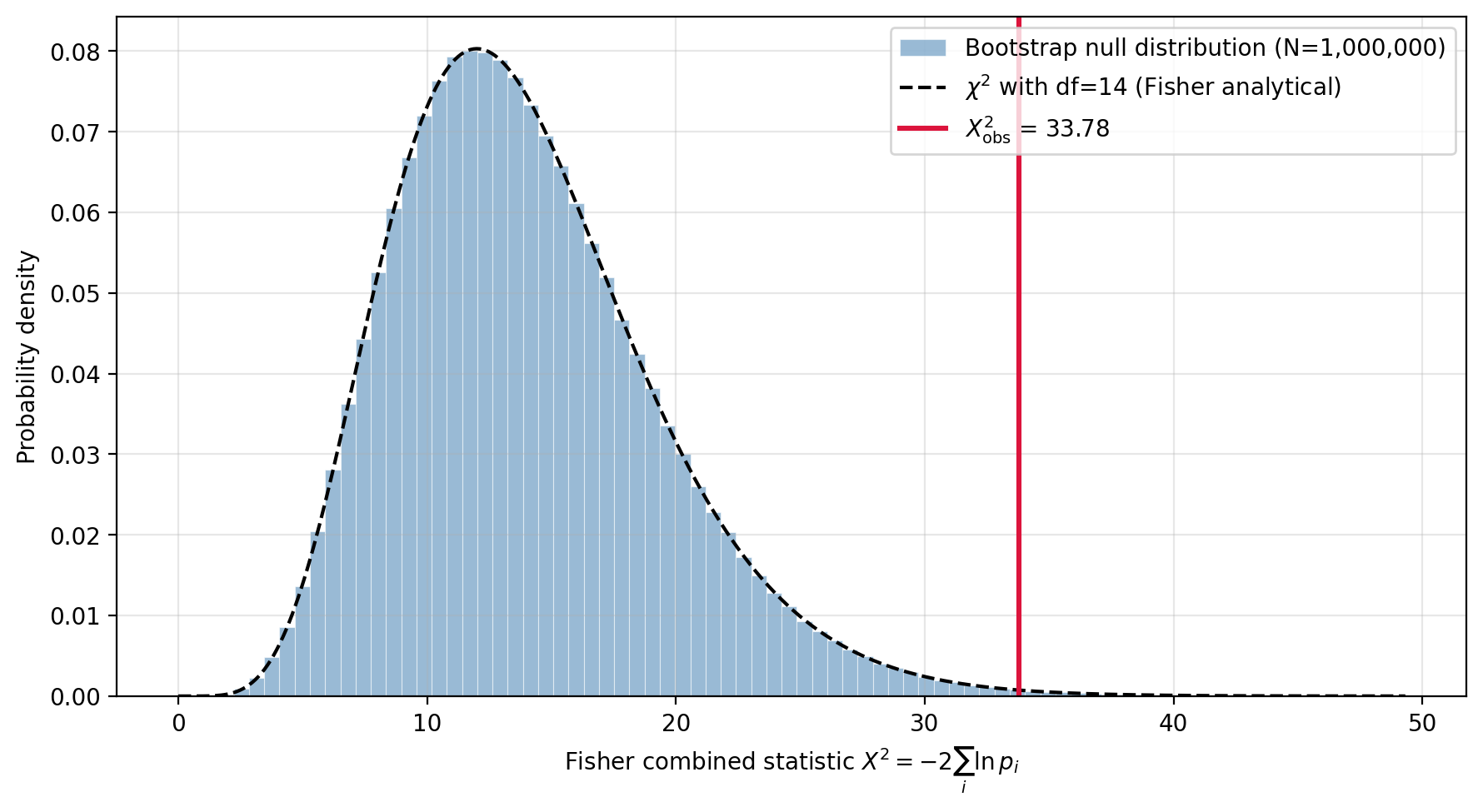}
    \caption{\textbf{Trials-corrected significance of the SPI detections.} 
Distribution of the Fisher combined statistic $X^2 = -2 \sum_i \ln p_i$, summed over 
all seven spectroscopic epochs. The blue histogram is the empirical null 
distribution from $10^6$ Monte Carlo trials, and the dashed black curve is the 
analytical $\chi^2$ distribution with $2N_{\rm epoch} = 14$ degrees of freedom. 
The red vertical line marks the observed value, $X^2_{\rm obs} = 33.78$, 
corresponding to a global p-value of $p_{\rm global} = 2.18 \times 10^{-3}$ .}
    \label{fig:Fischer}
\end{figure}

\begin{figure}
    \centering
    \includegraphics[width=0.8\linewidth]{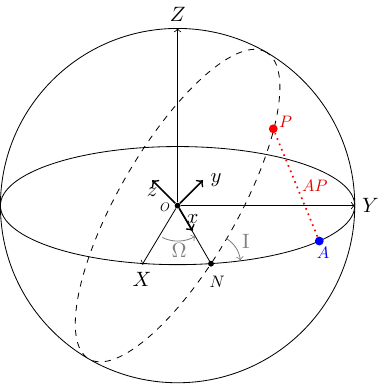}
    \caption{\textbf{Coordinate system and reference frame used in the geometric model.} The sphere represents the stellar surface, with the XYZ coordinate system centered on the origin (O), with the Z-axis towards the rotation axis. The stellar equator is contained in the XY plane, and the blue dot (A) represents the position of a stellar active region. The dashed circle is the projection of GJ~436~b orbit on the surface of the star. The orbit is contained in the xy plane and has an inclination \textit{I}. The black dot is the ascending node (N) and the x-axis points towards it. The red dot (P) is the position of an induced chromospheric hot spot that follows the planetary orbit. AP is the distance between the chromospheric hot spot induced by the planet (red) and the one intrinsic to the stellar activity (blue). $\Omega$ is the angle between the X and x axis. The red dotted line is the segment AP, the distance between the position of the stellar active region and the position of the induced chromospheric hot spot.}
    \label{fig:Coordinate}
\end{figure}

We identify additional peaks in the periodograms of various activity indicators, which we attribute to the rotation period, its half, or its double. Signals compatible with the rotation period are found in the $H\alpha$ (49.57~d), Na D (47.85~d) and Ca~{\sc ii} (49.57~d) activity indicators in H-08 (Fig. \ref{HARPS_2008_All}). The strongest signal comes from $H\alpha$, while the other two have FAP close or lower than 10\%. The H-09 season also exhibits signals in the rotation period range in $H\alpha$ (43.33~d) and in Ca~{\sc i} (41.11~d) (\ref{HARPS_2009_All}). The strongest peak in the Ca~{\sc ii} indicator is at twice the rotation period (88.93~d). Of the remaining epochs, we only find the rotation period in the $H\alpha$ of C-24 season (40.45~d), Fig.\ref{CARMENES_2024_All}. However, this peak has $7\%$ false alarm probability.

\paragraph{Ca~{\sc ii}~{\textsc{H\&K}} and Ca~{\sc ii}~{\textsc{IRT-a}} fluxes} \label{Sec: powers}
To estimate the planetary magnetic field from SPI, we must determine the power emitted in the SPI interaction from the spectroscopic observations. We use calibrations \cite{Labarga_2026} of the $\chi$ factor, the ratio between the continuum flux near a line and the bolometric flux \cite{Walkowicz_2004} for the CARMENES observations of the Ca~{\sc ii}~{\textsc{IRT-a}} line, because it is the only SPI signal with pEW measurements\\

The $\chi$ factor relates the fraction of emitted luminosity in a line ($L_{\text{line}}$) with respect to the bolometric luminosity ($L_{\text{bol}}$) to the equivalent width of the line: 

\begin{equation}
    \frac{L_{\text{line}}}{L_{\text{bol}}} = \chi \cdot {\rm pEW'}
    \label{eq:Lline_Lbol}
\end{equation}
\\

Using the calibrations \cite{Labarga_2022}, and the star's effective temperature ($T_{\rm eff}=3533\pm26$K, \cite{Marfil_2021}), we derive the $\chi$ factors for the Ca~{\sc ii}~{\textsc{IRT-a}} to be $7.99 \times 10^{-5}$. Then, we take the absolute value of the difference between the smallest and the largest values from the pEW' time series in Fig.~\ref{Fig_1}. We find $|\Delta {\rm pEW'}| = 0.0434 \, \pm \, 0.0274$ and $0.0308  \, \pm \, 0.0317$ for the C-16 and C-24 datasets, respectively. \\

From Eq.\ref{eq:Lline_Lbol}, we determine $L_{\rm line}$ using the Stefan-Boltzmann law for $L_{\rm bol} = 4\pi R^2 \sigma T_{\text{eff}}^4$. Given \(R = 0.422 \pm 0.01\,\) solar radii \cite{Maxted_2022} and \(T_{\text{eff}} = 3533\pm26\,\text{K}\), the emitted power of SPI in the Ca~{\sc ii}~{\textsc{IRT-a}} line is $(3.14\, \pm \, 2.0) \times 10^{19}\,\text{W}$ and $(2.23\, \pm \, 2.3) \times 10^{19}\,\text{W}$ in C-16 and C-24 observations, respectively.

\paragraph{Fraction of SPI detected power}\label{Sec:Frac_Power}
To determine the total energy of the SPI requires knowing the fraction dissipated in the lines where the emission is detected. SPI emission has been compared to that produced by solar flares \cite{Cauley_2019} or flares from other stars \cite{Loyd_2023}. Although this provides a first approximation, the real emission might be more like an auroral-like spot, as seen in the Jupiter-Io system \cite{Dulk_1965,Piddington_1968}. Despite having multiple spectral lines and activity indicators, we only detect a signal attributable to SPI in the Ca~{\sc ii}~{\textsc{H\&K}} and Ca~{\sc ii}~{\textsc{IRT-a}} lines. Therefore, it is likely that SPI emission is not as panchromatic as flare emission. Nevertheless, the estimation using flares serves as a lower limit for the fraction of energy emitted in these lines.\\

Following multiwavelength observations of AD Leonis \cite{Hawley_2003}, we find that the maximum emission in the Ca~{\sc ii}~{\textsc{IRT-a}} line is about 2\% of the total emission during the decay phase, with the Ca~{\sc ii}~{\textsc{IRT-a}} emission being half that. We compare the emission with the decay phase of the flare rather than the impulsive one, because by analogy with the Jupiter-Io system, we expect this type of interaction to be steady and longer-lasting, rather than short-lived and eruptive. In the case of AD Leonis, most of the energy radiated in optical lines is concentrated in H$\alpha$. Because we do not detect any signal in this line in our observations, we infer that the energy distribution of the SPI is less broad than in flares, where a larger fraction may be dissipated in the Ca~{\sc ii}~{\textsc{H\&K}} and Ca~{\sc ii}~{\textsc{IRT-a}} lines. A complementary explanation for the non-detection in the H$\alpha$ indicator could be that this line emits energy from the many small flares happening in the star \cite{Loyd_2023}. That would generate a higher background noise, making it harder to detect small variations caused by SPI. For our subsequent analyses, we assume the fraction could take any value ranging from 2 to 100\%.

\subsubsection*{Planetary magnetic field estimation} \label{Sec:SPI-models}

\paragraph{Energetics of different magnetic SPI models} 

We seek to estimate the planetary magnetic field strength by comparing the observed energy outputs from the SPI with theoretical predictions from various models. Analytical models have been proposed to quantify the power released through magnetic SPI mechanisms. We follow previous methods \cite{Cauley_2019} to obtain $B_p$ from different models, including Alfvén waves, magnetic reconnection, changes in the helicity, and an interconnecting loop between both bodies. All of these mechanisms give different frameworks to explain the interaction and account for the total power released. In the following sections, we summarize each model and compute the power available using the parameters of the GJ 436 system.

\paragraph{The Alfvén Wing Model}
This model \cite{Saur_2013}, is based on the Jupiter-Io interaction and does not require the planet to have an intrinsic magnetic field, but only a conducting atmosphere (ionosphere) or a conducting internal layer. The stellar wind in the vicinity of the planet is a magnetized plasma flow where the magnetic  field is frozen into the plasma. Therefore, perturbing the plasma flow, in particular slowing its velocity close to the surface of the planet that acts as an obstacle to the wind flow, produces a perturbation of the frozen in magnetic field of the wind. Such a perturbation is, in general, time dependent in the reference frame of the planet, even if the stellar wind flow is stationary at large distances from it. Therefore, the perturbation can be described as
the superposition of magnetohydrodynamic (MHD) waves. The fast mode of the perturbation propagate almost isotropically away from the planet, so the amplitude of those waves decreases rapidly with distance from the planet. The same happens, although over a larger distance, to the slow MHD mode. Only Alfvén waves reach the star because they remain focused propagating along the Alfvén wings with little dissipation \cite{Neubauer_1980,Saur_2013}.\\
In this scenario, magnetic power can propagate to the star only when the planet is inside the Alfvén surface of the stellar wind. The Alfvén surface is defined as the region of space where the Alfvén Mach number $M_{\rm A} \equiv v_{0}/v_{\rm A} < 1$, where $v_{0}$ is the velocity of the wind relative to the planetary body and $v_{\rm A} = B_{0} /\sqrt{\mu_{0} \rho}$ is the Alfvén velocity in the wind. In the latter term, $B_{0}$ is the intensity of the magnetic field in the wind, $\mu_{0}$ is the magnetic permeability of the vacuum, and $\rho$ is the wind plasma density. \\

The excited Alfvén waves move along the Alfvén characteristic lines [\cite{Saur_2013}, their section ~2.1.1]. To observe an effect in the stellar atmosphere, at least one of these characteristic lines must connect the planet with the star. The power $P_{\rm SPI}^{\rm AW}$ that reaches the star can be computed numerically. Assuming the Alfvén Mach number is small, we use the analytic approximation [\cite{Saur_2013}, their Eq.~(55)]

\begin{equation}
    P_{\rm SPI}^{\rm AW} = 2\pi R_{\rm eff}^{2} 
    \frac{ \left( \overline{\alpha} M_{\rm A} B_{0} \cos \Theta \right)^{2}}{\mu_{0}} v_{\rm A},
\end{equation}
where $0 < \overline{\alpha} <1$ is a parameter that measures the interaction strength between the planet and the wind, $R_{\rm eff}$ is the effective radius of the planet that acts as an obstacle to the wind flow, $\Theta$ is the angle between the velocity of the stellar wind flow and the normal to the magnetic field of the wind at the position of the planet. All the quantities ($M_{\rm A}, B_{0}, \Theta, v_{\rm A}$) are to be evaluated at the position of the planet. The effective radius $R_{\rm eff}$ is equal to the radius of the planet $R_p$ when it has no intrinsic magnetic field, otherwise it depends on the magnetic field of the planet and can reach a maximum value of $\sqrt{3} R_{\rm obst}$, where $R_{\rm obst} = R_p (B_p/B_{0})^{1/3}$ with $B_p$ the magnetic field at the equator of the planet [\cite{Saur_2013}], their section 2.3]. For a dipole field, the equatorial field is half the polar field. \\

The maximum power available for SPI in the Alfvén wing model is obtained for $R_{\rm eff} = \sqrt{3} R_p (B_p/B_{0})^{1/3}$, $\overline{\alpha}=1$, $\Theta = 0$ and corresponds to 
\begin{equation}
\begin{array}{lll}
P_{\rm SPI~(max)}^{\rm AW} & = & 6 \pi R_p^{2} (B_p/B_{0})^{2/3} M_{\rm A}^{2} B_{0}^{2} v_{\rm A}/\mu_{0} = \\
& = & (6\pi/\mu_{0}) M_{\rm A} R_p^{2} B_{0}^{4/3} B_p^{2/3} v_{0},
\end{array}
\label{p_aw_max}
\end{equation}
valid in the limit $M_{\rm A} \rightarrow 0$.\\

The strength of the interaction between the planet and the wind flow as parametrized by $\overline{\alpha}$ determines the excitation of the Alfvén waves propagating along the Alfvén wings. If the planet has no intrinsic magnetic field, the parameter $\overline{\alpha}$ depends on the electric conductivity of the planet solid body or of its ionosphere. The wind magnetic field $\vec{B}$ produces an induced electric field $\vec{E} = -\vec{v} \times \vec{B}$ because of the wind velocity $\vec{v}$ in the rest frame of the planet or of its ionosphere. The resulting current excites Alfvén waves with greater efficiency if the conductivity is high, compared to the case of a non-conducting planet for which the energy conveyed into the Alfvén wings is negligible. If the planet has an intrinsic magnetic field and a magnetosphere, the induced current propagates at the boundary of the magnetosphere where the plasma is strongly ionized and the efficiency of the alfvénic interaction becomes very high \cite{Saur_2013}.

\paragraph{The Magnetic Reconnection Model}

In this model, magnetic reconnection occurs between the stellar and planetary magnetic fields at the boundary of the planet's magnetosphere, which requires the planet to have an intrinsic magnetic field \cite{Lanza_2012}. The power released in this interaction is:

\begin{equation}\label{eq:Reconnection}
P^{\text{REC}}_{\text{SPI}} = \gamma \frac{\pi}{\mu_0} R^2 B_0^{4/3} B_p^{2/3} v_0,
\end{equation}
where $0 \leq \gamma \leq 1$ depends on the relative orientation of the magnetic field lines of the stellar and planetary fields. This equation accounts for the effective radius of the planetary magnetosphere. The value $\gamma \sim 1$ occurs when the magnetic fields have opposite directions, so reconnection has its maximum efficiency. Comparing Eqs.~(\ref{p_aw_max}) and~(\ref{eq:Reconnection}) 

\begin{equation}\label{p_ratio}
    \frac{P^{\text{AW}}_{\text{SPI}}(\text{max})}{P^{\text{REC}}_{\text{SPI}}(\text{max})} = 6 M_A < 1,
\end{equation}

in the limit in which Eq.~(\ref{p_aw_max}) is valid. Equation~\ref{p_ratio} shows that the power made available by the Alfv\'en wing mechanism is usually smaller than that delivered by the reconnection mechanism provided that $M_{\rm A} \ll 1$. To determine the order of magnitude of the available power, we could neglect the power conveyed by the Alfv\'en waves in comparison with that released by the reconnection. However, if the planet has no intrinsic magnetic field, $P_{\rm SPI}^{\rm REC} = 0$, so only the power conveyed by the Alfv\'en waves can produce an SPI signal in the stellar atmosphere.

\paragraph{The Helicity Variation Model}

Additional energy can be released in a tall stellar magnetic loop that extends up to the orbit of the planet \cite{Lanza_2012}. We take into account the variation of the relative magnetic helicity induced by the perturbations due to the orbital motion of the planet. The amount of available energy depends on the details of the stellar magnetic field configuration, which we assume to be force-free in the corona. The largest energy release is obtained when the field is a non-linear force-free field. A specific non-linear force-free field leads to available power 2 to 4 times larger than that released by the magnetic reconnection as modeled above [\cite{Lanza_2012}, their section 3.3]. Generalizing this result to a generic force-free field, we conclude that the helicity variation can provide an SPI power that is larger than that of the magnetic reconnection power, but of the same order of magnitude. Larger energy releases might occasionally be possible when a stellar loop perturbed by the planet has stored a large amount of magnetic energy and dissipates it by producing a strong flare. In such a case, a small perturbation by the orbiting planet can trigger the energy release, thus producing a preference for flaring when the planet passes over an active region that is ready to flare \cite{Lanza_2018}. 

\paragraph{The Interconnecting Loop Model}

This model assumes that the magnetic field of the stellar corona can reconnect with that of the planet and form a loop whose field lines interconnect the surface of the star with that of the planet, assuming that its atmosphere is completely ionized \cite{Lanza_2013}. The formation of such an interconnecting loop is possible only when the stellar field is potential (there are no electric currents flowing inside that region of the field) and remains in such a state during the subsequent energy dissipation phase. This is possible if the energy dissipation is very efficient inside the loop. In this model, energy is continuously accumulated inside the loop because of the stress applied by the orbital motion at the planetary footpoint and is immediately released to keep the configuration of the field potential. Because a potential field is the minimum-energy configuration for assigned boundary conditions on the star and the planet, such a strong dissipation makes the field stable by maintaining it in the potential configuration.\\

The fractional planetary surface area \( f_{AP} \) that acts as the footpoint of the interconnecting loop has been estimated from a model \cite{Adams_2011}, which depends on the ratio \( \zeta = B_0/B_p \) between the stellar magnetic field \( B_0 \) at the planet’s location and the planetary magnetic field \( B_p \):

\begin{equation} \label{f_ap}
f_{AP} = 1 - \sqrt{1 - \frac{3 \zeta^{1/3}}{2 + \zeta}} \quad .
\end{equation}\\

The power in the interconnecting loop model is then:

\begin{equation} \label{p_il}
P^{\text{IL}}_{\text{SPI}} = \frac{2 \pi}{\mu_0} f_{AP} R^2 B_p^2 v_{\text{orb}},
\end{equation}
where \( v_{\text{orb}} \) is the planet’s orbital velocity. This expression assumes that the stellar rotation period is much longer than the planet's orbital period, so the loop footpoint velocity is effectively equal to \( v_{\text{orb}} \). \( B_0 \) is indirectly included through \( \zeta \) in \( f_{AP} \). For an example planetary radius of $\text{R}_p \sim 10^8$ m, a field strength of $ \text{B}_p \sim \text{B}_0 \sim 10$ G, a velocity of $\text{v}_0 \sim 10^4$ to $10^5$ m\,$\text{s}^{-1}$ and $f_{AP} \sim 0.1$ to $0.2$, we calculate powers of $P^{\text{IL}}_{\text{SPI}}$\,$\sim 10^{20}$  to $10^{21}$ W.

\paragraph{Application to the GJ 436 system}\label{subsub:Bp}
We investigate which of the above SPI mechanisms could provide enough power to account for our observations. We focus our calculations on the excess power observed in C-16, because the planet is known to have been orbiting within the Alfvén surface at that time \cite{Vidotto_2023}. \\

We adopt stellar and planetary parameters, and a model of the stellar magnetic field, from previous work \cite{Vidotto_2023}. We compute the power available in the magnetic reconnection model and assume the following parameters in Eq.~\ref{eq:Reconnection}:
\begin{equation}
\begin{array}{ll}
 \gamma =1 \\ 
 R_p = 3. 85~R_{\oplus} = 24\,530~\mbox{km},\\
 B_{0} = 0.013~\mbox{G}, \\
 B_p = 1~\mbox{G}, \\
 v_{0} = 770~\mbox{km/s},    
\end{array}
\end{equation}
adopting the measured planet radius [\cite{Vidotto_2023}, their table 1], and a model for the magnetic field at the distance of the planet $B_{0}$ and the relative velocity between the planet and the wind $v_{0}$ [\cite{Vidotto_2023}, their Model I in their section 3]. We adopt the maximum value of $v_{0}$ in that model, to maximize the SPI power. We assess whether this model is capable of reproducing the observed powers; thus, the planetary magnetic field strength is assumed. The power available (Eq.~\ref{eq:Reconnection}) is $P_{\rm SPI}^{\rm REC} = 3.5 \times 10^{16}$~W. We regard this value as an upper limit, because we adopted the maximum value of $v_{0}$ from the wind model and $\gamma=1$. If we assume a stronger planetary field, the power available from magnetic reconnection scales with $B_p^{2/3}$. For example, a field of 10~G provides a power of $1.6 \times 10^{17}$~W. Matching the observations requires $B_p \sim 3.1 \times 10^{4}$~G, which we regard as implausibly high for a planet.\\

The Alfvén wing mechanism provides less power by a factor of $6\,M_{\rm A}$ (equation \ref{p_ratio}). In our case, adopting the same stellar wind model, the planet is close to the Alfvén surface where $M_{\rm A} = 1$, therefore Eq.~\ref{p_aw_max} might not be valid. However, that power is in the best case of the same order of magnitude as the reconnection power. Therefore, the Alfvén wing model also cannot account for the observation unless we assume an unreasonably strong planetary field. We reach a similar conclusion for the magnetic helicity variation model, because even in that case the expected gain with respect to the reconnection model is a factor $3-4$ assuming the same planetary magnetic field. \\

For the interconnecting loop model for $B_p$, the power obtained above for the C-16 observations is $3.14 \times 10^{19}\,\text{W}$. The orbital velocity, assuming a circular orbit, is $v_{\rm orb} = 2\pi a/P_{\rm orb} = 115.2\,km\,s^{-1}$. We consider two stellar magnetic fields at the distance of the planet of 0.013~G and 0.026~G [\cite{Vidotto_2023}, their Models I and II]. Substituting into Eq.~\ref{p_il} and solving for $B_p$, we find a minimum planetary magnetic field from the interconnecting loop model of $B_p = 10~\text{G}$ for the stellar magnetic field from Model I while $B_p = 9$~G for the one from Model II. \\

The stellar magnetic field at the distance of the planet affects the interconnecting loop model. We follow previous works \cite{Lanza_2018} and consider the most favorable case, a radial stellar field with an intensity $B_{\rm star}=27$~G at the poles of the star \cite{Bellotti_2023}. Such a field scales with the distance $r$ as $B_{0}(r)= B_{\rm star} (r/R)^{-2}$, where $R$ is the radius of the star. For GJ 436, the planet mean distance is $a/R= 14.56$ [\cite{Vidotto_2023}, their Table~1], leading to $B_{0}= 0.13$~G. The minimum planetary magnetic field in the interconnecting loop is then $B_p = 6.3$~G.\\

These values represent lower limits for the planetary magnetic field because we only observe a fraction of the total energy dissipated in the SPI event. Table \ref{Tab:Bfields} summarizes the $B_p$ values corresponding to different assumed fractions of the detected power for each model of the stellar magnetic field in the interconnecting loop scenario. \\
The full range for $B_p$ is 6.3~G to 110~G. The lower limit is consistent with theoretical studies of hot Neptunes around earlier-type stars \cite{Kilmetis_2024, Cauley_2019}.
Planetary magnetic fields between 6 and 15 Gauss are about one order of magnitude larger than the radial field components observed at the surface of Neptune and Uranus [\cite{Holme-Bloxham_1996}, their figures 10 to 14]. However, the stronger field of GJ 436~b could be accounted for by the larger mass of the planet or the strong stellar irradiation due to its proximity to its host star. A few percent of the stellar irradiation could be conveyed into the planetary interior \cite{Yadav_2017}, providing more energy for the operation of a planetary dynamo than available from the internal heat sources of the planet. In such a scenario, magnetic field strengths of $\sim 100~\text{G}$ are predicted for hot Jupiters, so we expect a few tens of Gauss for a Neptune-size planet such as GJ 436b.\\

The higher values in the range are likely overestimates of the true field strength. This broad range arises from uncertainties in both the fraction of energy dissipated in the detected spectral lines and the SPI model we apply.

\paragraph{Planetary magnetospheric radius} \label{subsub:Rm}
From the planet magnetic field estimates, we calculate the planetary magnetospheric radius $r_M$. This value refers to the front-facing radius of the magnetosphere --- the part pointing toward the star --- because it elongates in the opposite direction. $r_M$ depends on various factors, including as the magnetic field generating it ($B_p$), the surrounding environment, the orbital geometry and planetary tilt \cite{Pena-Monino_2024}. For the same $B_p$, if the planet is close to its star, its magnetosphere will be more compressed compared to a planet farther away, due to a combination of the thermal pressure of the plasma, the ram pressure of the stellar wind, and the star's magnetic pressure \cite{Ip_2004, Vidotto_2013}. If the external pressure is sufficiently high, the planet’s magnetosphere could be compressed to the point that part of its atmosphere becomes exposed to stellar wind erosion, potentially causing atmospheric loss. We adopt analytical functions to estimate the approximate magnetospheric radius \cite{Lanza_2009, Vidotto_2013, Lanza_2018}:

\begin{equation}\label{Eq:rM_Lanza}
    r_M = 2(2\delta_0)^{1/3} \left[ \frac{B_p}{B(\textbf{r}_p)} \right]^{1/3} R_p \quad .
\end{equation}
where $r_M$ is the magnetospheric radius, $\delta_0$ is a shape factor equal to 1.16 \cite{Lanza_2018}  and $\textbf{r}_p$ is the position vector of the planet. The scenario C in Table~\ref{Tab:Bfields} reports the magnetospheric radius of GJ~436~b $r_M = 9.6$ \text{to}  $21\,R_p$ using this model assuming different fractions of the total power emitted in the calcium lines.\\

These estimates are upper limits, because they primarily account for the stellar magnetic pressure, neglecting other pressures, such as ram pressure. Previous studies have characterized the stellar environment to study wind-planet interactions \cite{Bellotti_2023, Vidotto_2023}. From the equilibrium between the planetary magnetic pressure and the total stellar pressure, the magnetospheric radius \cite{Vidotto_2023}:

\begin{equation}\label{Eq:rM_Vidotto}
    r_M \simeq R_p \, \delta \left( \frac{(B_p/2)^2 /(8\pi)}{p_{\text{tot}}(a_{\text{orb}})} \right)^{1/6} \quad ,
\end{equation}
where $r_M$ is the magnetospheric radius, $\delta \simeq 2^{2/6}$ is a correction factor, $B_p$ is the planetary magnetic field, and $p_{\text{tot}}(a_{\text{orb}})$ is the average total pressure of the stellar wind at the planetary orbital distance [\cite{Vidotto_2023}, $8.2 \times 10^{-6} \, \text{dyn\,cm}^{-2}$ from their Model I and $59 \times 10^{-6} \, \text{dyn\,cm}^{-2}$ from Model II].\\

Previous work using a stellar magnetic field determination via spectropolarimetry \cite{Bellotti_2023} derived magnetospheric radii for GJ~436~b of $r_M = 5.2\,R_p$ and $r_M = 3.7\,R_p$ for those two models \cite{Vidotto_2023}. Those values were estimated assuming a planetary magnetic field of $B_p = 2\, \text{G}$ and an SPI power emission via Alfvén wings of $P_{\text{SPI}}^{\rm AW} = 1.2 \times 10^{15} \, \text{W}$ and $P_{\text{SPI}}^{\rm AW} = 4.8 \times 10^{15} \, \text{W}$, respectively. We excluded the Alfvén wing model above; we therefore re-calculate this value in the interconnecting loop scenario. Because we have detected the magnetic interaction between the planet and the star, we directly use our estimations of $B_p$ from Eq. \ref{f_ap} for several fractions of the detected SPI power. Substituting all the variables into Eq.~(\ref{Eq:rM_Vidotto}) we calculate $r_M$ for the various values of $B_p$ in Table~\ref{Tab:Bfields}. The results are shown in the scenarios A and B in Table~\ref{Tab:Bfields}, ranging from $r_M = 6 \, \, \text{to} \, \,20\,R_p$ for different percentages of the expected fraction of detected power and for the two wind models, consistent with a previous model \cite{Lanza_2018}.





\begin{table} 
	\centering
	\caption{\textbf{Extrapolated FAP obtained with the bootstrap windowing technique.} The FAP is listed for each identified SPI signal. These values were calculated with the bootstrap windowing technique for $N=5 \times 10^6$ random shuffles.}
	\label{FAP-table} 
	
	\begin{tabular}{cccc} 
		\\
		\hline
		Dataset & Period [d] & ln(FAP) & FAP\\
		\hline
		H-08 & 2.81 & -5.59  & $3.7 \times 10^{-3}$ \\
		C-16 & 2.81 & -4.87 & $7.7 \times 10^{-3}$ \\
		C-24 & 2.46 & -8.06 & $3.2 \times 10^{-4}$ \\
		\hline
	\end{tabular}
\end{table}

\begin{table}[h]
    \centering
    \caption{\textbf{Per-epoch p-values from the unified bootstrap procedure and their contributions to the Fisher combined statistic.} 
    Per-epoch p-values $p_i$ correspond to the probability that random noise would produce a peak in 
    the search range $[2.36, 3.0]$~d as strong as observed.}
    \label{table:fisher}
    \begin{tabular}{lcc}
    \\
        \hline
        Epoch & $p_i$ & $-2\ln p_i$ \\
        \hline
        H-07 & 0.0720 & 5.26 \\
        H-08 & 0.0210 & 7.73 \\
        H-09 & 0.8734 & 0.27 \\
        H-10 & 0.9200 & 0.17 \\
        H-20 & 0.2642 & 2.66 \\
        C-16 & 0.0645 & 5.48 \\
        C-24 & 0.0022 & 12.23 \\
        \hline
        \textbf{Total ($X^2_{\rm obs}$)} & & \textbf{33.78} \\
        \hline
    \end{tabular}
\end{table}


\clearpage 



\end{document}